\documentclass[aps,twocolumn,showpacs,preprintnumbers,nofootinbib,floatfix,prd,superscriptaddress,10pt]{revtex4-2}
\usepackage[colorlinks=true,linktocpage=true,linkcolor=blue,citecolor=red,urlcolor=blue]{hyperref}
\usepackage{graphicx,amssymb,amsmath,amsthm,amsfonts,epsfig,epsf,fixmath}
\usepackage{aas_macros}
\usepackage[usenames]{color}
\usepackage{epstopdf}
\usepackage{bbold}
\usepackage{bm}
\usepackage{comment}
\usepackage{dcolumn}
\usepackage{enumitem}
\usepackage{latexsym}
\usepackage{rotating}
\usepackage{enumerate}
\usepackage{tensor,multirow}
\usepackage{url}
\usepackage{ragged2e}

\newcommand{\bea}{\begin{eqnarray}}
\newcommand{\eea}{\end{eqnarray}}

\begin{document}

\title{Impact of Neutrino Trapping on the Equation of State and Structure of Hybrid Stars with 2SC Quark Matter}
\author{Andrea Sabatucci}
\email{andrea.sabatucci@uwr.edu.pl}
\affiliation{Institute of Theoretical Physics, University of Wroc\l{}aw,
50-204 Wroc\l{}aw, Poland}

\author{Armen Sedrakian}
\email{armen.sedrakian@uwr.edu.pl, sedrakian@fias.uni-frankfurt.de}
\affiliation{Institute of Theoretical Physics, University of Wroc\l{}aw,
50-204 Wroc\l{}aw, Poland}
\affiliation{Frankfurt Institute for Advanced Studies, D-60438
Frankfurt am Main, Germany}
  \begin{abstract}
    We investigate how sensitive the effects of neutrino trapping on
    hot hybrid stars are to uncertainties in the hadronic and quark
    equations of state and in the thermodynamic construction of the
    hadron--quark transition.  The hadronic phase is described by
    finite-temperature covariant density functional equations of
    state, and the quark phase by a three-flavor Nambu--Jona-Lasinio
    (NJL) model with repulsive vector interactions and two-flavor
    color superconductivity.  Because confinement is absent from the
    NJL model, an effective pressure shift $B^*$ is used
    phenomenologically to locate the deconfinement transition.  We
    construct isentropic stellar equations of state with and without
    trapped electron neutrinos, vary the quark vector coupling and the
    nuclear symmetry-energy slope, and compare global with local
    conservation of electron-lepton number within phase coexistence.
    Neutrino trapping generically shifts deconfinement toward higher
    density throughout the parameter range considered.  Increasing the
    vector interaction stiffens quark matter and raises the maximum
    stellar mass, but only moderately for the couplings studied.  We
    also examine stronger diquark pairing and RG-consistent
    regularization, instead of sharp momentum cut-off: both affect the
    pairing gap, phase boundaries, and quark-core composition more
    strongly than they alter the qualitative trends of the stellar
    mass--radius sequences.  Conversely, trapping reduces the
    sensitivity to the hadronic symmetry-energy slope because the
    enhanced proton fraction drives matter toward greater isospin
    symmetry.  Global and local lepton-number prescriptions produce
    rather similar mass--radius relations but substantially different
    internal phase structures: global conservation leads to a
    considerably more extended mixed phase.  
\end{abstract}

\maketitle

\section{Introduction}
\label{sec:intro}

The possible appearance of deconfined quark matter in compact stars
remains one of the central questions in the physics of strongly
interacting matter at high density.  While cold neutron stars probe
matter in approximate beta equilibrium at temperatures small compared
with the relevant Fermi energies, newly born in a type-II supernova
proto-neutron stars and binary neutron-star merger remnants explore a
qualitatively different thermodynamic regime~(for reviews
see~\cite{Oertel2017RvMP,Baiotti2019,Sedrakian2026}).  In these
systems, temperatures may reach tens of MeV, and neutrinos can remain
trapped on timescales long compared with the local equilibration
time~\cite{Espino2024,Pajkos2025}.  The matter is therefore
characterized not only by its baryon density, but also by a nonzero
entropy per baryon and a finite lepton fraction.  These additional
thermodynamic constraints can modify both the location and the nature
of a transition between hadronic and quark matter.  

Recent microscopic calculations provide additional support for
treating neutrinos as trapped also after the onset of
color-superconducting quark matter.  In particular,
Ref.~\cite{Hofmann2025} evaluated neutrino absorption mean
free paths in the 2SC phase within an NJL model for densities and
temperatures relevant to neutron-star mergers.  For equilibrated,
lepton-rich 2SC matter, the absorption mean free paths of degenerate
neutrinos were found to be of order meters or less, i.e., much smaller
than the characteristic length scale of a compact-star core.  These
results indicate that neutrino trapping can persist in the
color-superconducting quark phase and therefore provide microscopic
motivation for treating electron-lepton number as a conserved quantity
across the hadron--quark transition.

The treatment of the hadron--quark transition under such conditions is
particularly subtle.  In neutrino-transparent matter, electric charge
is the only relevant conserved charge in addition to baryon number.
A first-order phase transition between locally charge-neutral phases is
usually implemented by a Maxwell construction when baryon number is
the only independent conserved charge.  In matter with
trapped electron neutrinos, however, the electron-lepton number is
conserved on the relevant dynamical timescale.  If electric neutrality
is imposed locally while lepton number is conserved globally, the two
coexisting phases can carry different lepton fractions.  The
coexistence conditions then involve both the baryon and lepton
chemical potentials, and the pressure varies continuously through the
mixed phase.  This form of phase equilibrium was discussed in
Refs.~\cite{Hempel-Pagliara,Pagliara-PRL} and provides a natural
framework for studying hot hybrid stars with trapped neutrinos.

Several mutually complementary approaches have been used to
investigate quark deconfinement in proto-neutron stars and hot compact
stars.  Calculations based on nonlocal chiral quark models have
demonstrated that temperature and color superconductivity can
substantially modify hybrid-star sequences and, in some cases,
generate thermally induced twin-star configurations~\cite{Carlomagno2024,Carlomagno_Blaschke}. A recent study based on a confining density functional approach suggested that the emergence of thermal twin-star branches could provide an indicator of the explodability of massive blue supergiant stars in core-collapse supernovae~\cite{blaschke_iva_2026}. Proto-neutron-star matter
has also been studied using nonlocal Polyakov--NJL models~\cite{Malfatti2019}, statistical mean-field descriptions matched to
NJL quark matter \cite{Ghaemmaghami2023}, and alternative quark-matter
models along evolutionary sequences~\cite{Mariani2017}.  More
recently, renormalization-group-consistent formulations of the NJL
model have been developed in order to improve its behavior at large
chemical potentials~\cite{Gholami_RGfirst,Gholami2025,RGNJL_stars,kunkel2026}.  These studies
illustrate that predictions for hot hybrid stars can depend not only
on temperature and lepton content, but also on the microscopic
description of both phases and on the prescription adopted for phase
equilibrium.

In Ref.~\cite{Sabatucci2026}, we investigated this problem using a
finite-temperature covariant density-functional equation of state for
the hadronic phase and a three-flavor NJL model with
vector interactions and two-flavor color-superconducting (2SC)
pairing for quark matter~\cite{Alford2008}.
The analysis was performed along isentropic
trajectories with and without trapped electron neutrinos.  For the
neutrino-trapped case, global conservation of electron-lepton number
was imposed while each phase was required to be locally electrically
neutral.  The resulting coexistence region was therefore constructed
using the generalized Gibbs conditions appropriate to two globally
conserved charges.  Within a representative choice of the hadronic
and quark-matter parameters, it was found that neutrino trapping
shifts deconfinement toward higher baryon densities, substantially
changes the particle composition, and modifies the masses and radii
of hot hybrid-star configurations.  In particular, the additional
thermal and leptonic pressure leads to configurations that are less
compact than their cold counterparts.  The thermodynamic construction
used in Ref.~\cite{Sabatucci2026} will serve as the reference
framework for the present study.

The purpose of the present work differs from Ref.~\cite{Sabatucci2026}.  Rather than repeating
the baseline analysis, we examine how robust its conclusions are when
the principal model assumptions are varied. There are several sources of
uncertainty that are especially relevant. On the quark-matter side,
the repulsive vector interaction controls the stiffness of the
high-density equation of state and therefore affects both the location
of the phase transition and the maximum mass of hybrid configurations.
The phenomenological ``deconfinement'' pressure shift $B^*$, which is
introduced in the NJL description to control the relative position of
the hadronic and quark pressures, provides an additional uncertainty
in the transition density.  On the hadronic side, the density
dependence of the nuclear symmetry energy affects the properties of
neutron-rich matter and hence the matching to the quark phase.
Finally, in neutrino-trapped matter there is a conceptual uncertainty
associated with the treatment of lepton-number conservation: one may
conserve electron-lepton number globally, as in the Gibbs construction
adopted in Ref.~\cite{Sabatucci2026}, or impose it locally in each
phase, which leads to a Maxwell-type construction~\cite{Bonanno2012}.

We therefore extend the analysis of Ref.~\cite{Sabatucci2026} in
several directions.  First, we study explicitly how the transition
changes with the pressure shift $B^*$ and use the pressure difference
between the two phases to characterize the range in which a
first-order transition persists at finite temperature.  Second, we
vary the strength of the quark vector interaction through
$\eta_V=G_V/G_S$ and determine its effect on the isentropic equation
of state and on stellar masses and radii.  Third, we repeat the
calculation for members of the DDLS family with different values of
the symmetry-energy slope $L_{\rm sym}$~\cite{Li_2023} in order to assess the
sensitivity to the hadronic isovector sector.  This comparison is
particularly useful in neutrino-trapped matter, because the large
electron and proton fractions drive the hadronic phase toward greater
isospin symmetry and can consequently reduce its sensitivity to
$L_{\rm sym}$.  We additionally examine a stronger diquark coupling,
$\eta_D=1.25$, and quantify the changes obtained when conventional
cutoff regularization is replaced by an RG-consistent NJL scheme.
Finally, we compare the generalized Gibbs construction
with a Maxwell construction at fixed lepton fraction.  We examine not
only their effect on bulk stellar observables, but also the resulting
radial distribution of the hadronic, mixed, and quark phases inside
massive stars.

An important point of the latter comparison is that agreement at the
level of the mass--radius relation does not necessarily imply similar
internal stellar structures.  Two phase-transition prescriptions may
produce only modest differences in global observables while predicting
substantially different spatial extensions of the mixed phase.  Such
differences can become relevant for quantities that depend directly
on the microscopic composition of the stellar core, including weak
interaction rates, neutrino transport, thermal conductivity,
viscosities, and phase-conversion dynamics.  The distinction is
therefore potentially more important for the dynamical evolution of a
proto-neutron star or merger remnant than would be inferred from
static mass--radius sequences alone.

Our analysis consequently separates two questions that are sometimes
not clearly separated.  The first is whether the qualitative effect of neutrino
trapping on quark deconfinement is robust.  The second is how accurately
the location and spatial extent of the phase transition, and the
associated stellar properties, can be predicted in the presence of
uncertainties in the hadronic and quark equations of state and in the
treatment of phase coexistence.  By varying these ingredients within
a common thermodynamic framework, we identify which features of hot
2SC hybrid stars are generic and which remain quantitatively
model-dependent.

The paper is organized as follows.  In
Sec.~\ref{sec:baryonic_eos} we summarize the finite-temperature
hadronic equations of state and specify the range of hadronic models
considered here.  Section~\ref{sec:quark_eos} introduces the 2SC NJL
description and discusses the parameters of the quark sector that are
varied in this work.  The equilibrium conditions for the
hadron--quark transition in neutrino-trapped and neutrino-transparent
matter are reviewed in Sec.~\ref{sec:PT_trapped_neutrinos}.  Our
numerical results are presented in Sec.~\ref{sec:results}, including
the dependence on the vector interaction and hadronic symmetry energy,
the effects of the diquark coupling and RG-consistent regularization,
and the comparison between Gibbs and Maxwell constructions.
We summarize our conclusions in Sec.~\ref{sec:discussion}.

\section{Hadronic Matter}
\label{sec:baryonic_eos}

For the hadronic phase we employ the finite-temperature equations of
state constructed within covariant density functional theory in
Ref.~\cite{Tsiopelas2024EPJA}.  These equations of state cover the
ranges of baryon density, temperature, and charge fraction relevant
for proto-neutron stars and binary neutron-star merger remnants and
are available in tabulated form through the CompOSE database.  In the
present work we restrict the hadronic composition to nucleons and
leptons.  A detailed description of the thermodynamic construction
and its implementation in the neutrino-trapped hybrid-star problem
was given in Ref.~\cite{Sabatucci2026}; here we summarize only those
features needed for the model comparison carried out below.

The density-dependent covariant functionals belong to the DDLS family
\cite{Li_2023}.  Their isoscalar properties are kept fixed, while the
density dependence of the isovector sector is varied through the
slope of the nuclear symmetry energy,
\begin{equation}
 L_{\rm sym}
 =
 3 \varrho_0
 \left.
 \frac{\partial E_{\rm sym}(\varrho)}{\partial \varrho}
 \right|_{\varrho=\varrho_0},
\end{equation}
where $\varrho_0$ denotes the saturation density.   We consider the three
representative values
\begin{equation}
\label{eq:L_sym_values}  
 L_{\rm sym}=30,\;50,\;70~{\rm MeV},
\end{equation}
while keeping the remaining parameters of the DDLS construction,
in particular the skewness parameter
$Q_{\rm sat}=400~{\rm MeV}$, fixed.  The $L_{\rm sym}=50$ MeV model
coincides with the baseline hadronic equation of state adopted in
Ref.~\cite{Sabatucci2026}.  The additional $L_{\rm sym}=30$ and
$70$ MeV cases allow us to quantify how uncertainties in the
isovector sector propagate to the deconfinement transition and to
the properties of hot hybrid stars.

This comparison is particularly relevant because the sensitivity to
the symmetry energy depends strongly on the composition of the
matter.  In neutrino-transparent $\beta$-equilibrated matter, the
large neutron excess makes the high-density equation of state
sensitive to the density dependence of the symmetry energy.  Trapped
electron neutrinos modify this balance.  At a fixed electron-lepton
fraction $Y_{Le}$, the electron abundance is increased and charge
neutrality consequently requires a larger proton fraction.  The
hadronic phase is therefore driven toward a more isospin-symmetric
composition, reducing the influence of the symmetry-energy slope.
One of the aims of the present work is to determine to what extent
this tendency suppresses the spread among the different DDLS models
in the resulting stellar sequences.

For neutrino-trapped matter we impose $\beta$ equilibrium according to
\begin{equation}
 \mu_n-\mu_p=\mu_e-\mu_{\nu_e},
\end{equation}
together with local electric charge neutrality.  Equivalently, using
the chemical potential $\mu_{Le}$ associated with electron-lepton
number,
\begin{equation}
 \mu_{\nu_e}=\mu_{Le}.
\end{equation}
The finite-temperature DDLS tables provide the baryonic thermodynamic
quantities as functions of density, temperature, and charge fraction,
while the neutrino contribution to the pressure, energy density, and
entropy is added assuming a non-interacting ultrarelativistic Fermi
gas.  The global constraint on $Y_{Le}$ and its role in constructing
the hadron--quark coexistence region are discussed in
Sec.~\ref{sec:PT_trapped_neutrinos}.

At sub-saturation densities the high-density tables are joined to the
corresponding nonuniform-matter equation of state used for the stellar
calculations, following the same prescription as in
Ref.~\cite{Sabatucci2026}.  Since the focus here is the sensitivity of
the high-density transition and stellar core to the hadronic
isovector sector, this low-density treatment is kept unchanged for
all members of the DDLS family.

\section{2SC Quark Matter}
\label{sec:quark_eos}

The high-density phase is described by the same three-flavor NJL model
employed in Ref.~\cite{Sabatucci2026}.  The model contains
scalar--pseudoscalar, vector, and diquark interaction channels,
supplemented by the six-fermion 't~Hooft interaction that breaks the
axial $U_A(1)$ symmetry.  We retain the microscopic parametrization of
the previous work and focus here on the two ingredients that are most
relevant for the model dependence of the hadron--quark transition: the
strength of the repulsive vector interaction and the phenomenological
pressure shift $B^*$.

Neglecting electromagnetic interactions, the effective Lagrangian is
\begin{eqnarray}
\label{eq:NJL-Lagrangian}
\mathcal{L}_{\rm NJL}
&=&\bar{\psi}(i\gamma^\mu\partial_\mu-\hat m)\psi\nonumber\\
&& \hspace{-1cm}+G_S\sum_{\alpha=0}^{8}
\left[(\bar{\psi}\lambda^{\alpha}\psi)^2+(\bar{\psi}i\gamma_5\lambda^{\alpha}\psi)^2\right]
\nonumber\\
&&\hspace{-1cm}-G_V(\bar{\psi}\gamma^\mu\psi)^2\nonumber\\
&&\hspace{-1cm}+G_D\sum_{\gamma,c}\left[\bar{\psi}^{a}_{\alpha}i\gamma_5
\epsilon^{\alpha\beta\gamma}\epsilon_{abc}(\psi_C)^b_\beta\right]
\left[(\bar{\psi}_C)^r_\rho i\gamma_5\epsilon^{\rho\sigma\gamma}\epsilon_{rsc}\psi^s_\sigma\right]
\nonumber\\
&&\hspace{-1cm}-K\left\{\det_f[\bar{\psi}(1+\gamma_5)\psi]+
\det_f[\bar{\psi}(1-\gamma_5)\psi]
\right\},
\end{eqnarray}
where $\psi_\alpha^a$ denotes a quark field with flavor
$\alpha=u,d,s$ and color $a=r,g,b$, and
$\hat m=\mathrm{diag}_f(m_u,m_d,m_s)$.

We use $ m_{u,d}=5.5~{\rm MeV},$ $m_s=140.7~{\rm MeV},$
$\Lambda=602.3~{\rm MeV},$ together with $G_S\Lambda^2=1.835,$
$K\Lambda^5=12.36 ,$ as in the standard three-flavor NJL
parametrization of Ref.~\cite{RehbergPhysRevC}.  Our reference results
use $G_D=G_S$~\cite{Sabatucci2026}, and we also consider the stronger
pairing interaction $G_D=1.25G_S$.  We denote the corresponding
coupling ratios by $\eta_D=G_D/G_S$ and $\eta_V=G_V/G_S$.  These
variations enable us to separate the influence of diquark pairing from
that of repulsive vector interactions.

At the densities and lepton content relevant for the transition considered here, the
paired state is the 2SC phase.  Up and down quarks of two colors form
Cooper pairs in the color-antitriplet channel, whereas strange quarks
remain unpaired.  The corresponding condensate can be written
schematically as
\begin{equation}
\Delta_c
\propto
G_D
\left\langle
(\bar{\psi}_C)^a_\alpha i\gamma_5
\epsilon^{\alpha\beta c}\epsilon_{abc}
\psi^b_\beta
\right\rangle .
\end{equation}
Chiral symmetry breaking generates the constituent masses
\begin{equation}
M_\alpha
=
m_\alpha
-4G_S\sigma_\alpha
+2K\sigma_\beta\sigma_\gamma ,
\end{equation}
with $(\alpha,\beta,\gamma)$ cyclic in flavor space.  The vector mean
fields shift the effective quark chemical potentials, while electric
and color neutrality are enforced through the corresponding chemical
potentials.  The condensates and neutrality chemical potentials are
determined self-consistently from the stationarity conditions of the
thermodynamic potential.

For completeness, the pressure in the mean-field approximation is
written as
\begin{widetext}
\begin{eqnarray}
\label{eq:Pressure}
P
&=&
\frac{1}{2\pi^2}
\sum_{i=1}^{18}
\int_0^\Lambda dk\,k^2
\left[
|\epsilon_i|
+
2T\ln\left(1+e^{-|\epsilon_i|/T}\right)
\right]
+4K\sigma_u\sigma_d\sigma_s
\nonumber\\
&-&
\frac{1}{4G_D}\sum_{c=1}^{3}|\Delta_c|^2
-2G_S\sum_{\alpha=1}^{3}\sigma_\alpha^2
+\frac{1}{4G_V}(2\omega_0^2+\phi_0^2)
+P_e
-P_0-B^* ,
\end{eqnarray}
\end{widetext}
where $\epsilon_i$ are the quasiparticle dispersion relations and
$P_0$ fixes the vacuum normalization.  The displayed expression
contains the quark and electron terms.  In neutrino-trapped matter
the ideal-gas contribution $P_{\nu_e}$ is added to the total
pressure and its energy and entropy contributions are included in
the equation of state; the muon-lepton sector is neglected throughout.
The pressure shift $B^*$ is independent of the chemical potentials and
temperature, and hence does not itself contribute to the number or
entropy densities.

\subsection{Vector interaction and effective deconfinement shift}

The principal microscopic parameter varied in the quark sector is the
strength of the repulsive vector interaction, which we characterize by
\begin{equation}
\label{eq:eta_def}
\eta_V=\frac{G_V}{G_S}.
\end{equation}
The reference calculation of Ref.~\cite{Sabatucci2026} employed
$\eta_V=1$.  In the present work we consider the interval
\begin{equation}
\label{eq:eta_interval}
0.8\leq\eta_V\leq1.2 ,
\end{equation}
in order to quantify the sensitivity of the hybrid equation of state
and stellar configurations to the stiffness of quark matter.

The vector interaction modifies the density dependence of the quark
pressure itself.  Increasing $\eta_V$ enhances the repulsion between
quarks and generally stiffens the high-density equation of state.
It therefore affects both the thermodynamic competition between the
hadronic and quark phases and the maximum masses supported by hybrid
configurations.  One of the questions addressed below is whether this
microscopic uncertainty changes the qualitative consequences of
neutrino trapping or mainly shifts the corresponding transition
densities and stellar observables.

In addition to the interaction parameters of the NJL model, we
introduce a constant pressure offset $B^*$ through
Eq.~(\ref{eq:Pressure}).  Its physical status is different from that
of $G_S$, $G_D$, and $G_V$.  The NJL model describes spontaneous
chiral symmetry breaking and color superconductivity, but it does not
contain a confinement mechanism.  Consequently, the relative
normalization of the pressures obtained from the hadronic and NJL
descriptions is not fixed by a microscopic description of the
deconfinement transition.  The parameter $B^*$ is introduced
phenomenologically to compensate for this limitation.

In this sense, $B^*$ should be regarded as an effective
deconfinement parameter rather than as an additional microscopic
interaction.  It shifts the quark-matter pressure according to
$
P_Q\longrightarrow P_Q-B^*,
$
and thereby controls the chemical potential or density at which quark
matter becomes thermodynamically favored over hadronic matter.  Along
a given thermodynamic trajectory, the hadronic phase is favored as
long as
$
P_H>P_Q ,
$
whereas beyond the crossing the quark phase becomes favored when
$
P_Q>P_H .
$
The equality
$
P_H=P_Q
$
therefore marks the coexistence condition, supplemented, of course,
by the appropriate chemical and thermal equilibrium constraints
discussed in Sec.~\ref{sec:PT_trapped_neutrinos}.

For the reference case $\eta_V=1$ we adopt
\begin{equation}
\label{eq:B_star_value}
B^*=-20~{\rm MeV\,fm^{-3}},
\end{equation}
which yields an onset of deconfinement at approximately
$\varrho_B\simeq3.5\,\varrho_0$ at zero temperature for the reference
hadronic equation of state~\cite{Sabatucci2026}.  This choice is not meant to identify a
fundamental vacuum-energy difference between the two phases.
Instead, it fixes the effective location of the deconfinement
transition within the hybrid construction.

\begin{figure*}
\includegraphics[width=.8\textwidth]{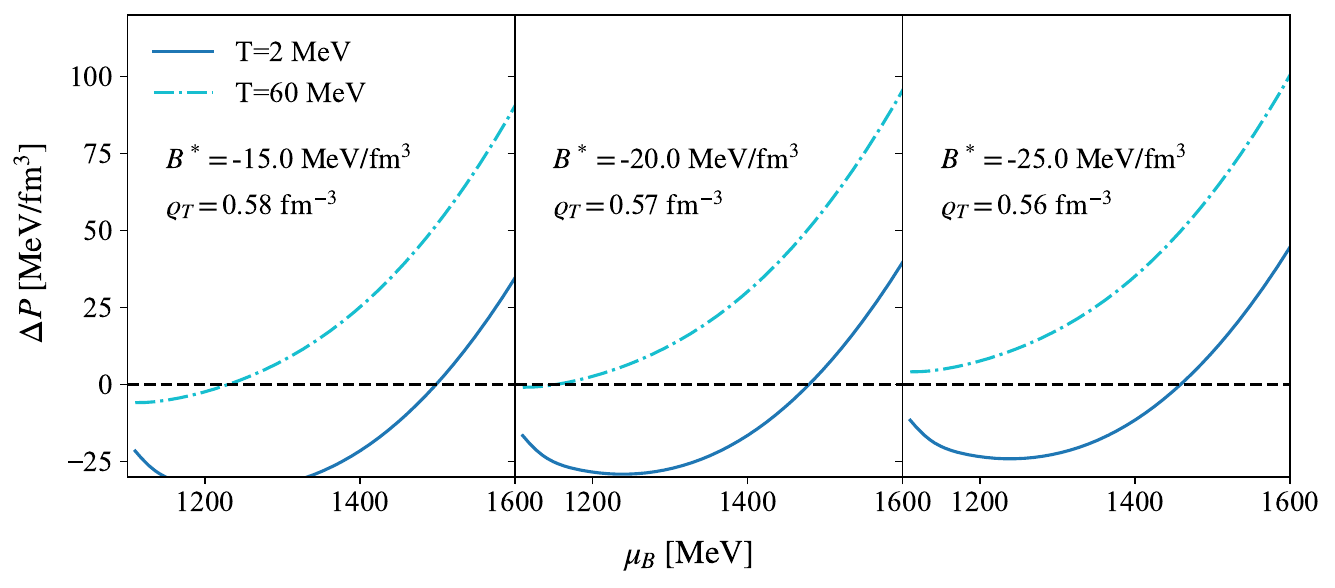}
\caption{
  Difference $\Delta P=P_Q-P_H$ between the quark- and
  hadronic-matter pressures as a function of the baryon chemical
  potential for three values of the effective pressure shift
  $B^*$.  The solid and dashed curves are evaluated for neutrino-transparent matter at $T=2$ and
  $60$~MeV, respectively, as indicated in the plot.
  Regions with $\Delta P<0$ favor hadronic matter, whereas
  $\Delta P>0$ favors quark matter. The zeros of $\Delta P$ identify
  the mechanical coexistence points and illustrate how the
  phenomenological parameter $B^*$ controls the effective location of
  the deconfinement transition.  With the convention
  $P_Q=P_Q^{(0)}-B^*$, a more negative $B^*$ raises $P_Q$ and
  shifts its crossing with the hadronic pressure.}\label{fig:Bag}
\end{figure*}

The influence of this prescription is illustrated in
Fig.~\ref{fig:Bag}, where we show
\begin{equation}
\Delta P=P_Q-P_H
\end{equation}
as a function of the baryon chemical potential for several choices
of $B^*$ in neutrino-transparent matter.  Negative values of $\Delta P$ correspond to a region in
which the hadronic phase has the larger pressure and is
thermodynamically preferred.  At $\Delta P=0$ the two phases satisfy
the mechanical coexistence condition, while $\Delta P>0$ indicates
that quark matter has become favored.  The figure therefore shows
directly how the effective pressure offset controls the location of
the hadron--quark crossing.

Comparing the $T=2$ and $T=60$ MeV curves additionally reveals how
the competition between the two phases evolves with temperature.
The latter temperature is used here to illustrate the pressure
competition beyond the temperatures sampled by the stellar isentropes.
Our parameter choices are restricted to cases in which a
first-order hadron--quark transition remains present over the
temperature range relevant for the isentropic configurations
considered below.  This allows us to investigate the effects of
neutrino trapping, vector repulsion, and hadronic-model dependence
without mixing them with a disappearance of the first-order
transition itself.

\section{Phase Equilibrium and Lepton-Number Conservation}
\label{sec:PT_trapped_neutrinos}

At temperatures of several MeV and above, the neutrino mean free path
in dense stellar matter can become sufficiently short that electron
neutrinos are trapped on the dynamical timescales relevant for
proto-neutron stars and binary neutron-star merger remnants
\cite{Hofmann2025}.  In this regime the electron-lepton
number must be included among the conserved quantities that determine
phase equilibrium.  This changes the thermodynamics of a first-order
hadron--quark transition relative to cold, neutrino-transparent
matter.

The treatment adopted in Ref.~\cite{Sabatucci2026}, and used as our
reference here, is along Refs.~\cite{Hempel-Pagliara,Pagliara-PRL}.
Electric charge neutrality is imposed locally in each phase, whereas
baryon number and electron-lepton number are conserved globally.
The latter assumption allows the two coexisting phases to have
different local lepton fractions while sharing a common lepton
chemical potential.  The purpose of the present work is not only to
use this construction, but also to compare it systematically with a
Maxwell-type treatment in which the lepton fraction is constrained
locally.

\subsection{Global lepton-number conservation}

We first consider the generalized Gibbs construction appropriate when
electron-lepton number can be redistributed between the coexisting
hadronic and quark phases.  At a fixed temperature, the thermodynamic
state of each locally charge-neutral phase may be written as
\begin{equation}
P=P(T,\mu_B,\mu_{Le}),
\end{equation}
where $\mu_B$ and $\mu_{Le}$ are the chemical potentials associated
with baryon number and electron-lepton number, respectively.

In the hadronic phase, beta equilibrium requires
\begin{equation}
\mu_n=\mu_p+\mu_e-\mu_{\nu_e},
\end{equation}
and local charge neutrality gives
\begin{equation}
Y_p=Y_e.
\end{equation}
Introducing the electric-charge chemical potential $\mu_Q$, the
particle chemical potentials can be expressed as
\begin{equation}
\begin{split}
\mu_n &= \mu_B,\\
\mu_p &= \mu_B+\mu_Q,\\
\mu_e &= -\mu_Q+\mu_{Le},\\
\mu_{\nu_e} &= \mu_{Le}.
\end{split}
\end{equation}
For quark matter,
\begin{equation}
\mu_u=
\frac{1}{3}\mu_B+\frac{2}{3}\mu_Q,
\qquad
\mu_d=\mu_s=
\frac{1}{3}\mu_B-\frac{1}{3}\mu_Q ,
\end{equation}
supplemented by the color-neutrality conditions of the 2SC phase.

Coexistence between the locally neutral hadronic and quark phases
requires thermal, mechanical, and chemical equilibrium with respect
to the globally conserved charges,
\begin{equation}
\begin{split}
T^H &=T^Q,\\
\mu_B^H&=\mu_B^Q,\\
\mu_{Le}^H&=\mu_{Le}^Q,\\
P_H(T,\mu_B,\mu_{Le})
&=
P_Q(T,\mu_B,\mu_{Le}).
\end{split}
\label{eq:Gibbs_conditions}
\end{equation}
At fixed temperature these conditions define a coexistence line in
the $(\mu_B,\mu_{Le})$ plane.  The construction is therefore
qualitatively different from the one-dimensional pressure crossing
encountered in cold neutrino-free matter.

For astrophysical applications we follow trajectories at a prescribed
total electron-lepton fraction
\begin{equation}
Y_{Le}=\frac{\varrho_e+\varrho_{\nu_e}}{\varrho_B}.
\end{equation}
The numerical comparisons for trapped matter use $Y_{Le}=0.4$,
representative of lepton-rich proto-neutron-star conditions.  In each
phase, $\varrho_{Le}=\varrho_e+\varrho_{\nu_e}$ includes both charged leptons and trapped
electron neutrinos.  Inside the
coexistence region the quark volume fraction $\chi$ determines the
average baryon density,
\begin{equation}
\bar \varrho_B
=
(1-\chi)\varrho_B^H+\chi \varrho_B^Q,
\end{equation}
while global conservation of electron-lepton number gives
\begin{equation}
Y_{Le}
=
\frac{
 (1-\chi)\varrho_{Le}^{H}
 +\chi \varrho_{Le}^{Q}}
{\bar \varrho_B}.
\label{eq:global_YLe}
\end{equation}
The corresponding energy and entropy densities are
\begin{equation}
\bar\epsilon
=
(1-\chi)\epsilon_H+\chi\epsilon_Q,
\end{equation}
and
\begin{equation}
\bar s
=
(1-\chi)s_H+\chi s_Q.
\end{equation}

An important consequence of Eq.~(\ref{eq:global_YLe}) is that the
local lepton fractions of the two phases need not be equal.  During
the transition, electron-lepton number can be redistributed between
hadronic and quark matter while the prescribed total $Y_{Le}$ is
maintained.  The coexistence pressure therefore changes continuously
as $\chi$ evolves from zero to unity.  The mixed phase consequently
occupies a finite interval in density even at fixed temperature,
unlike the standard Maxwell construction with a single globally
conserved charge.  This was one of the central features of the
reference calculation in Ref.~\cite{Sabatucci2026}.

\subsection{Local lepton-number constraint: Maxwell benchmark}
\label{sec:maxwell_formalism}

To assess how strongly the results depend on this treatment of
lepton-number conservation, we also construct an alternative
Maxwell-type equation of state.  In this case the prescribed lepton
fraction is imposed separately in each phase rather than only on the
mixed system as a whole.  Electron-lepton number is therefore treated
as a local constraint and cannot be redistributed freely between the
hadronic and quark components.

For matter at fixed $Y_{Le}$, the thermodynamic potential relevant
for phase coexistence is the Gibbs free energy per baryon.  The
coexistence condition can be written as
\begin{equation}
\mu_B^H+Y_{Le}\mu_{Le}^H
=
\mu_B^Q+Y_{Le}\mu_{Le}^Q ,
\label{eq:maxwell_condition}
\end{equation}
together with equality of the temperature and pressure.  In the
neutrino-free limit, where the lepton chemical potential ceases to
represent an independent conserved quantity, this reduces to the
usual Maxwell condition involving equality of the baryon chemical
potential.

The distinction between Eqs.~(\ref{eq:Gibbs_conditions}) and
(\ref{eq:maxwell_condition}) is physical rather than merely
technical.  In the generalized Gibbs construction, lepton number is
shared globally between the phases and $\mu_{Le}$ is common across
the interface.  In the Maxwell treatment, the same value of
$Y_{Le}$ is imposed locally on each phase and the corresponding
lepton chemical potentials need not coincide.  The two constructions
therefore describe different assumptions about how efficiently
lepton number can rearrange during phase conversion.

We use the Maxwell construction here primarily as a controlled
benchmark.  It allows us to determine which observables are
insensitive to the detailed implementation of phase equilibrium and
which retain information about the internal composition of the mixed
region.  As shown below, the distinction can be modest at the level
of global stellar mass--radius relations while becoming considerably
more pronounced in the radial phase structure of massive hybrid
stars.

\subsection{Isentropic trajectories}

Proto-neutron stars and merger remnants are more naturally
characterized by entropy per baryon than by a globally fixed
temperature.  We therefore construct stellar equations of state along
trajectories satisfying
\begin{equation}
S=\frac{s}{\varrho_B k_B}=\mathrm{const}.
\end{equation}
As in Ref.~\cite{Sabatucci2026}, the calculation is based on a set of
thermodynamically consistent isothermal equations of state covering
temperatures between $2$ and $60$~MeV.  
The isentropic trajectories are ob
tained by interpolating between
these isothermal solutions while imposing the entropy constraint
explicitly at each density.

For the Gibbs construction, an isentrope enters and leaves the
coexistence surface at generally different temperatures.  Within the
mixed phase, the temperature, pressure, and quark volume fraction
therefore evolve simultaneously so as to satisfy both the fixed
entropy and fixed total lepton-fraction constraints.  This procedure
produces a smooth isentropic equation of state through the
coexistence region.

A finite coexistence interval can also arise along an isentropic
trajectory in the Maxwell construction.  This does not contradict the
familiar constant-pressure plateau of an isothermal Maxwell
transition.  At fixed entropy per baryon the two phases generally
carry different entropy densities, so that maintaining a constant
total $S$ requires both the temperature and the phase fraction to
change as the system crosses the transition.  Consequently, the
pressure may vary along an isentropic coexistence trajectory whereas
it remains constant for the corresponding isothermal Maxwell
construction.

The same thermodynamic treatment is applied to neutrino-transparent
matter, for which the independent electron-lepton constraint is
removed ($\mu_{\nu_e}=0$, without fixing $Y_{Le}$) and the usual
Maxwell construction is recovered.  Comparing
trapped and neutrino-free isentropes then isolates the effects caused
by lepton trapping from those arising from finite entropy alone.

\subsection{Low-density matching and stellar configurations}

For the calculation of stellar configurations, the high-density
equations of state are complemented at
$\varrho_B\lesssim0.5\,\varrho_0$ by the corresponding nonuniform-matter
description based on the HS(DD2) model
\cite{HEMPEL2010210}, following the prescription used in
Ref.~\cite{Sabatucci2026}.  The same low-density treatment is adopted
for all parameter sets considered here so that the comparisons below
reflect changes in the high-density hadronic and quark sectors rather
than differences in the stellar envelope.

The resulting isentropic equations of state are used to solve the
Tolman--Oppenheimer--Volkoff equations
\cite{Tolman,OppVolkoff}.  Throughout the present study we neglect the
muon-lepton sector and concentrate on the consequences of trapped
electron neutrinos.  This keeps the phase-equilibrium problem confined
to the two globally conserved quantities, baryon number and
electron-lepton number, in the Gibbs construction.

The radial stability of the reference isentropic hybrid-star sequences
was examined by Rau, Sabatucci, and Sedrakian~\cite{Rau2026}.
For the equilibrium mixed-phase equations of state studied there,
including the globally constrained Gibbs and locally constrained
Maxwell-type constructions, the fundamental radial mode becomes marginal
($\omega_0^2=0$) at the maximum-mass turning point and unstable beyond it.
The radial-mode stability boundary therefore agrees with the
Bardeen--Thorne--Meltzer criterion for these sequences.  This result
provides an independent stability check of the reference models,
although the additional coupling and regularization variations
considered below were not individually subjected to a radial-mode
analysis.

\section{Results}
\label{sec:results}

We now examine the robustness of the finite-temperature hybrid-star
construction against the variations introduced in the preceding
sections.  The reference model corresponds to the setup studied in
Ref.~\cite{Sabatucci2026}: the DDLS equation of state with
$L_{\rm sym}=50$~MeV and $Q_{\rm sat}=400$~MeV is used for the
hadronic phase, while the quark sector is described by the 2SC NJL
model with $\eta_V=1$ and the corresponding effective deconfinement
pressure shift $B^*$ given by Eq.~\eqref{eq:B_star_value}.

The thermodynamic properties of this reference model, including the
finite-temperature phase diagram, particle composition, and
mass--radius sequences, were discussed in detail in
Ref.~\cite{Sabatucci2026}.  In particular, that work showed that
trapped electron neutrinos shift the onset of deconfinement toward
higher density, modify the charge composition of both phases, and
increase the radii of hot stellar configurations.  Rather than
reproducing those results, we use them here as a reference point and
focus on the additional information obtained by varying the
microscopic and thermodynamic assumptions.

\subsection{Phase boundary and neutrino trapping}
\label{sec:results_baseline}

\begin{figure}[tp]
    \includegraphics[width=.9\linewidth]{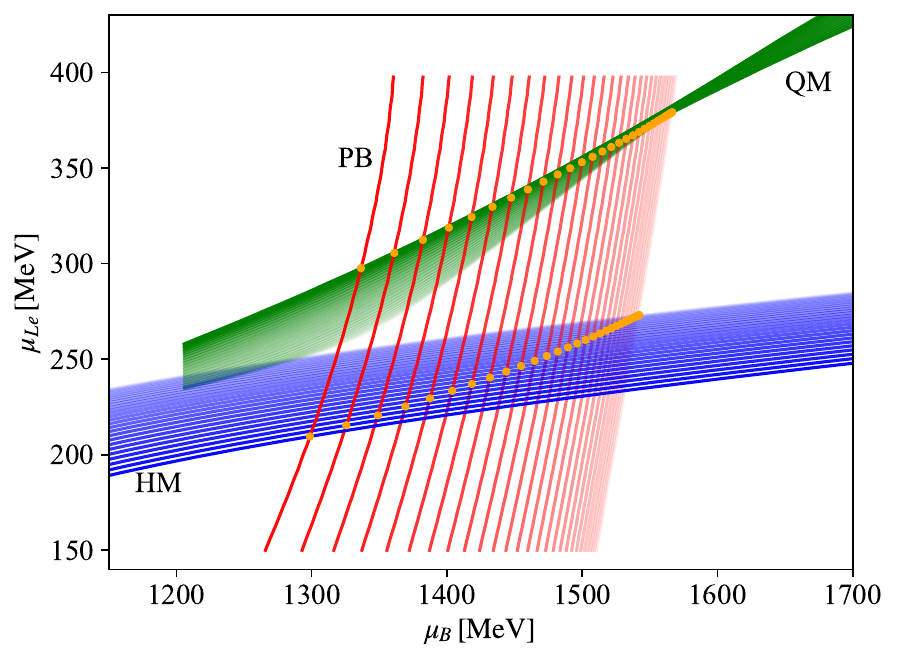}
    \caption{
    Projection of the phase-equilibrium construction onto the
    $(\mu_B,\mu_{Le})$ plane for the reference model at
    $Y_{Le}=0.4$.  The trajectories corresponding to locally
    charge-neutral hadronic and quark matter are shown together with
    the coexistence line obtained from the generalized Gibbs
    conditions.  The filled circles indicate the entrance into and
    exit from the mixed phase. For each phase, lighter shades correspond to lower temperatures.
    }
    \label{fig:chempot_proj}
  \end{figure}

The geometry of the phase construction is particularly transparent
in the $(\mu_B,\mu_{Le})$ plane.  Figure~\ref{fig:chempot_proj}
shows the trajectories of locally charge-neutral hadronic and quark
matter at fixed total electron-lepton fraction, together with the
coexistence line determined by the Gibbs conditions of
Sec.~\ref{sec:PT_trapped_neutrinos}, for several temperatures in the interval $2-60\,\rm MeV$.  The two intersections of the
fixed-$Y_{Le}$ trajectories with the coexistence line identify,
respectively, the entrance into and exit from the mixed phase.

This representation emphasizes a feature that is obscured in a
one-dimensional Maxwell construction.  Because electron-lepton number
is conserved globally, the coexistence condition is not specified by
a single crossing in $\mu_B$.  Instead, the system evolves along a
line in the two-dimensional space of conserved-charge chemical
potentials while lepton number is redistributed between the two
phases.  The mixed phase therefore extends over a finite range of
thermodynamic states.

\begin{figure}[tp]
    \centering
    \includegraphics[width=\linewidth]{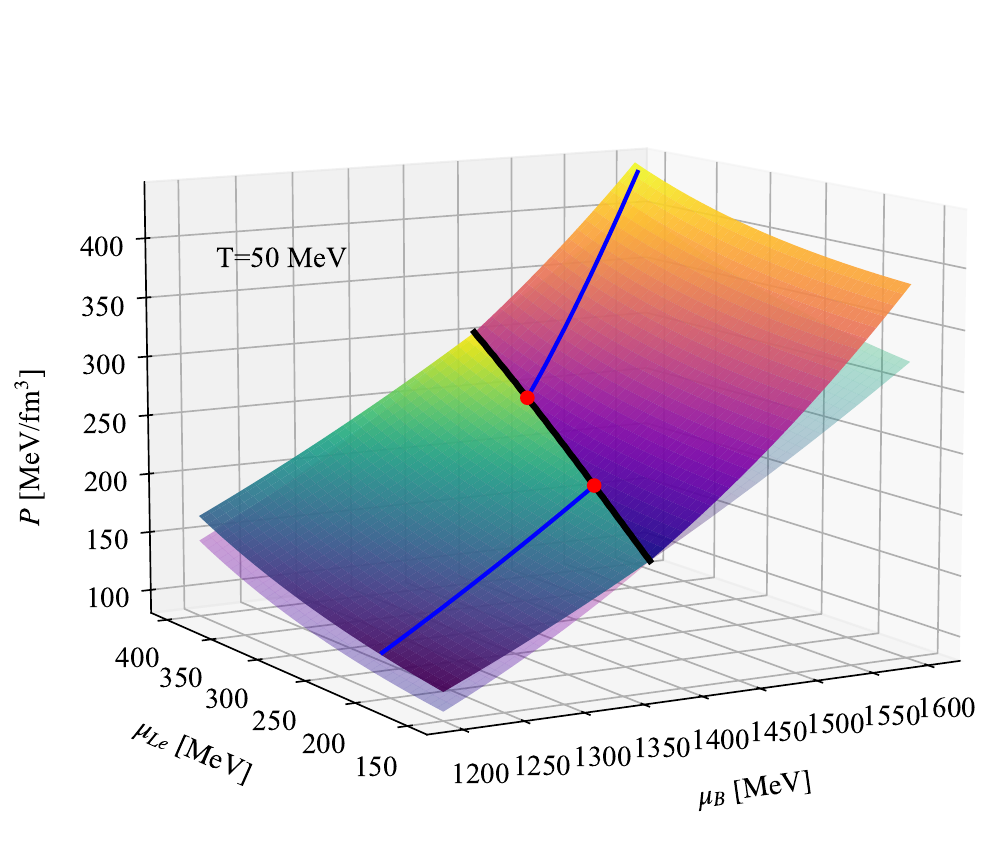}
    \caption{Hadronic and quark pressure surfaces at $T=50$~MeV for
    the reference model ($\eta_V=\eta_D=1$), displayed against the
    baryon and electron-lepton chemical potentials.  Their
    intersection traces the phase-coexistence curve.  The plot
    complements its two-dimensional projection in
    Fig.~\ref{fig:chempot_proj}.}
    \label{fig:3Dplot}
\end{figure}

The geometry can also be seen in the pressure surfaces at fixed
temperature shown in Fig.~\ref{fig:3Dplot}.  Here the coexistence
curve is the intersection of the hadronic and quark pressure surfaces
in the space $(\mu_B,\mu_{Le},P)$.  Its projection onto the
$(\mu_B,\mu_{Le})$ plane is the coexistence line discussed above.
This representation is useful because the chemical potential associated
with trapped lepton number is an independent thermodynamic variable.

\begin{figure}[tp]
    \includegraphics[scale=0.5]{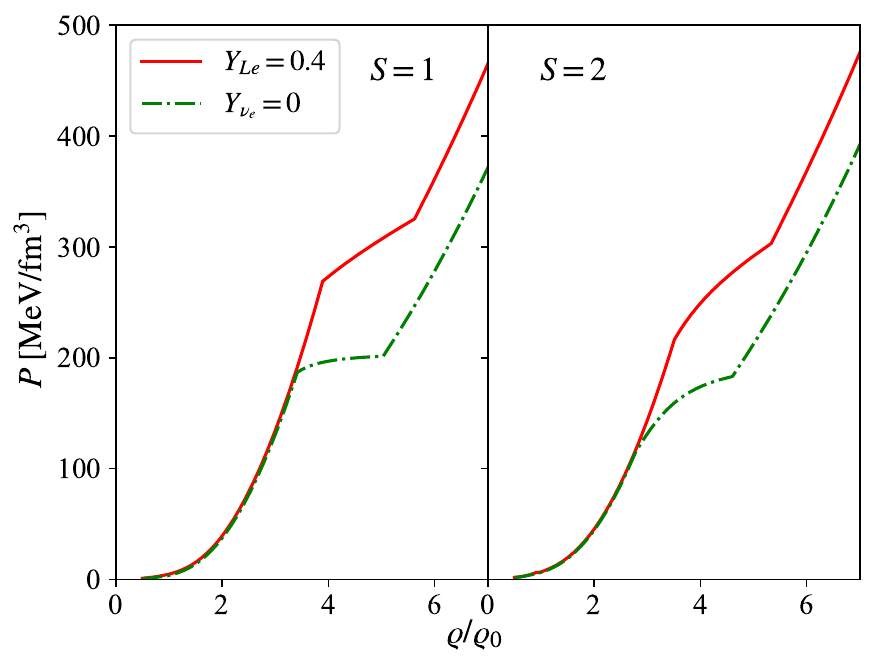}
    \caption{
    Comparison of isentropic pressure--density relations for
    neutrino-transparent matter and matter with trapped electron
    neutrinos.  Neutrino trapping shifts the onset of the
    hadron--quark transition toward higher baryon density.  The
    remaining model parameters are those of the reference
    configuration.
    }
    \label{fig:Iso_ye_comp}
  \end{figure}

A more direct view of the impact of neutrino trapping is provided by
Fig.~\ref{fig:Iso_ye_comp}, which compares the isentropic
pressure--density relations obtained with and without trapped
electron neutrinos.  The most prominent effect is the displacement of
the hadron--quark transition toward higher baryon density in
lepton-rich matter.  This confirms that the delay of deconfinement
found in Ref.~\cite{Sabatucci2026} is not tied to a particular
representation of the phase diagram, but is directly visible in the
stellar equation of state.

The origin of the shift is rooted in the composition of matter.  Fixing a nonzero
$Y_{Le}$ requires a substantial electron abundance and, through local
charge neutrality, a correspondingly enhanced proton fraction in the
hadronic phase.  The resulting matter is less neutron rich and
remains thermodynamically favored relative to quark matter over a
larger density interval.  Deconfinement therefore occurs only after
a larger compression has been reached.

  \subsection{Dependence on the quark vector interaction}
\label{sec:results_vector}

\begin{figure}[tp]
    \includegraphics[width=\linewidth]{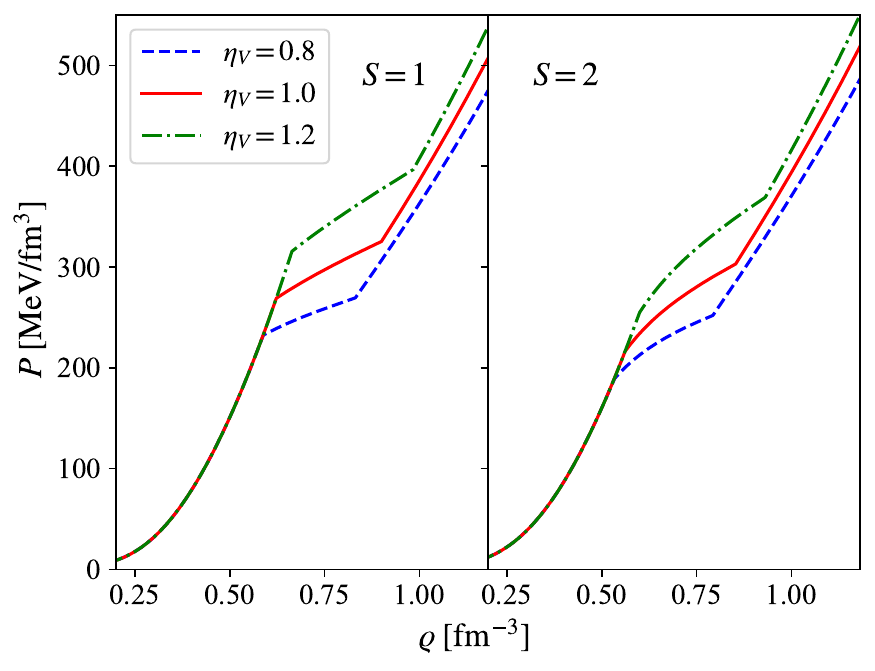}
    \caption{
    Isentropic pressure--density relations for different strengths
    $\eta_V=G_V/G_S$ of the repulsive vector interaction in the
    quark phase.  Increasing $\eta_V$ stiffens the high-density
    equation of state, while leaving the qualitative behavior of the
    hadron--quark transition unchanged.
    }
    \label{fig:gv_iso_comp}
\end{figure}

We next examine the uncertainty associated with the repulsive vector
interaction in the quark sector.  Figure~\ref{fig:gv_iso_comp}
compares the isentropic equations of state obtained for several
values of $\eta_V$ in the interval \eqref{eq:eta_interval}.

The expected stiffening of quark matter with increasing $\eta_V$ is
clearly reflected in the pressure--density relation.  This effect is
primarily associated with the quark-containing part of the equation
of state: the hadronic sector is unchanged, whereas the larger
repulsive vector mean field increases the pressure supported by the
high-density phase.  The transition itself can also be displaced
because the relative hadronic and quark pressures entering the
coexistence condition are modified. To disentangle effects of the microscopic parameters and the choice of $B^*$ on the location of the phase transition, we adopt the same prescription for selecting $B^*$ in all models considered: {\it $B^*$ is chosen such that the transition occurs at the
  lowest density compatible with the persistence of a first-order
  transition in neutrino-transparent matter up to $T=60$ MeV.} Detailed information about the values of $B^*$ and the associated onset densities $\varrho_{\rm tr}$ can be found in the Appendix~\ref{chapter:tables}.
  
Importantly, however, varying the vector interaction over the range
considered here does not change the qualitative thermodynamic picture.
In particular, the characteristic effect of neutrino trapping, namely
the shift of deconfinement toward larger density, persists throughout
the parameter range considered.  Thus, the transition density is
model-dependent, whereas the direction of the trapping effect remains
the same across this set of calculations.

\begin{figure}[tp]
    \includegraphics[width=\linewidth]{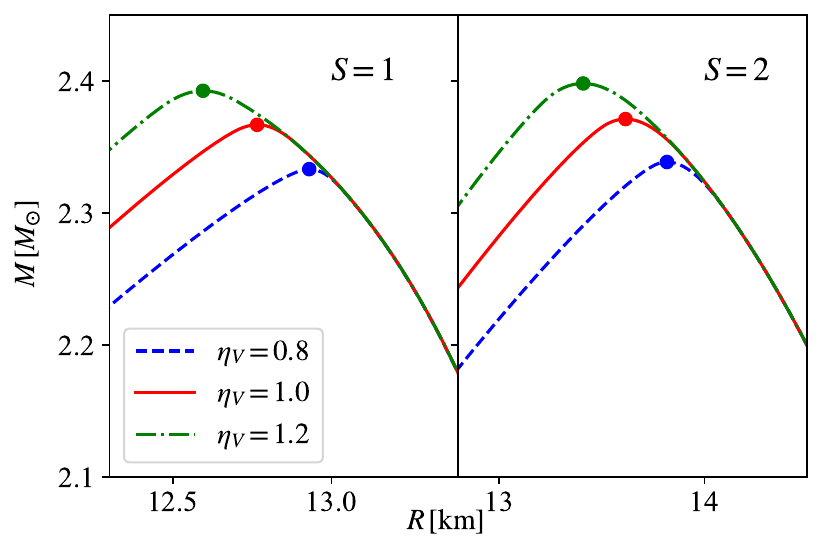}
    \caption{
    Mass--radius relations for isentropic hybrid stars obtained for
    different values of the vector coupling $\eta_V$.  Increasing
    the repulsive vector interaction stiffens the quark-matter
    equation of state and produces a moderate increase of the
    maximum stellar mass.}
    \label{fig:MR_gv_comp}
  \end{figure}

The corresponding mass--radius sequences are shown in
Fig.~\ref{fig:MR_gv_comp}.  Increasing $\eta_V$ increases the maximum
mass, as expected from the additional repulsion in the quark phase.
For $0.8\leq\eta_V\leq1.2$, however, the change in the maximum mass
is only at the level of a few percent.  The stellar calculation
therefore leads to a useful separation between qualitative and
quantitative model dependence: the vector coupling affects the
absolute high-density stiffness and hence the mass scale, but it does
not qualitatively alter the consequences of finite entropy and
neutrino trapping.

This result also indicates that, within the range explored here,
uncertainty in the NJL vector coupling is unlikely by itself to erase
the characteristic differences between neutrino-rich and
neutrino-transparent stellar sequences.  It should instead be viewed
mainly as an uncertainty in the quantitative location of the phase
transition and in the maximum mass supported by the quark core.
A fixed-$B^*$ comparison that isolates the vector-coupling variation
is presented in Sec.~\ref{sec:GD_discussion}, together with the
quantitative maximum-mass comparison in Table~\ref{tab:masses}.

  \subsection{Dependence on the hadronic isovector sector}
\label{sec:results_Lsym}

We now turn to the uncertainty associated with the hadronic equation
of state.  The calculation is repeated for the DDLS models with values
$L_{\rm sym}=30$, $50$, and $70$ MeV given in
Eq.~\eqref{eq:L_sym_values}. The isoscalar parameters and other model settings are kept fixed.

In making this comparison, care is required because the effective
pressure shift $B^*$ can be used to move the hadron--quark crossing
independently of the intrinsic differences among the hadronic models.
We therefore apply the same prescription discussed in Sec.~\ref{sec:results_vector} to each member of the DDLS family. 
With this criterion, the onset densities are approximately $\varrho_{\rm tr}\simeq0.57~{\rm fm}^{-3}$ for all the hadronic configurations considered.

This prescription is useful because it prevents the comparison from being dominated by an arbitrary relative displacement of the hadronic and NJL pressures.  The residual differences between the stellar sequences can then be interpreted primarily in terms of the density dependence of the hadronic equation of state.

\begin{figure*}
    \includegraphics[width=0.7\linewidth]{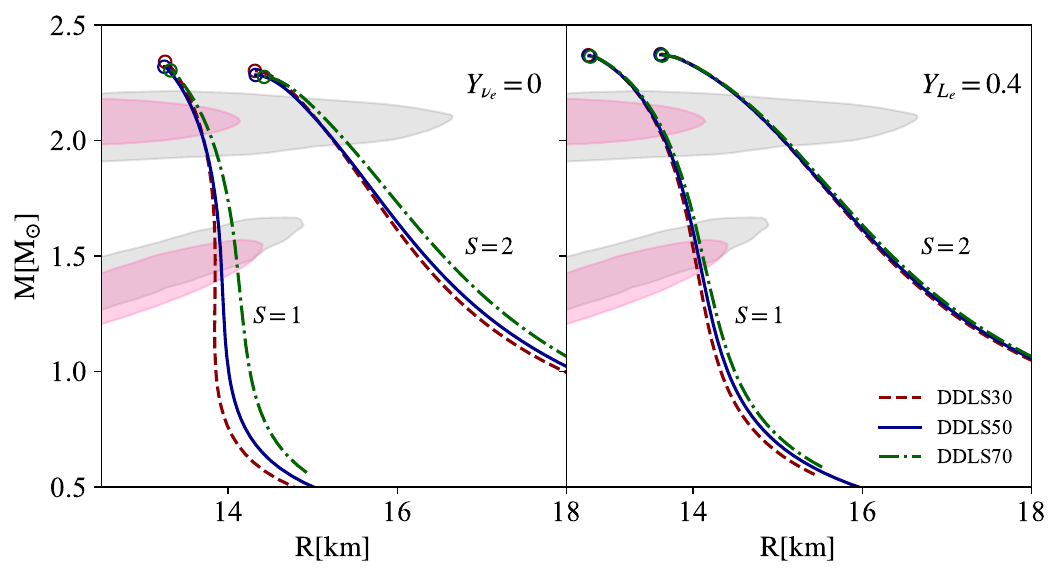}
    \caption{
    Mass--radius relations for isentropic hybrid stars obtained with
    the DDLS hadronic equations of state having different symmetry-energy
    slopes $L_{\rm sym}$.  The left panel shows neutrino-transparent
    configurations, while the right panel shows neutrino-trapped
    configurations.  The reduced spread of the curves in the trapped
    case reflects the smaller isospin asymmetry of lepton-rich matter.
    }
    \label{fig:MR_DDLS_comp}
  \end{figure*}

The resulting mass--radius relations are displayed in
Fig.~\ref{fig:MR_DDLS_comp}.  In neutrino-transparent matter the
different values of $L_{\rm sym}$ produce a visible spread among the
stellar sequences.  This is expected because cold beta-equilibrated
matter is strongly neutron rich and is therefore sensitive to the
isovector part of the nuclear interaction.

The situation changes noticeably when neutrinos are trapped.  The
spread among the different DDLS models is reduced.  The physical
origin of this behavior can be understood directly from the
composition.  At finite $Y_{Le}$ the larger electron abundance
requires a larger proton fraction, reducing the isospin asymmetry
$ \delta=1-2Y_p .  $ Since the leading symmetry-energy contribution to
the energy per particle is approximately proportional to
$ E_{\rm sym}(\varrho)\,\delta^2, $ the sensitivity to the density
dependence of $E_{\rm sym}$, and therefore to $L_{\rm sym}$, is
suppressed as $\delta$ decreases.

The convergence of the trapped-neutrino sequences is therefore not an
accidental feature of the selected parameter sets.  It reflects a
general tendency of lepton-rich matter toward greater isospin
symmetry.  {\it Neutrino trapping consequently reduces an important
  source of hadronic equation-of-state uncertainty}, as is seen for
the stellar configurations considered here.  The effects of trapping
on the radii and maximum masses remain qualitatively the same for all
three DDLS models.

  \subsection{Global versus local lepton-number conservation}
\label{sec:results_gibbs_maxwell}

\begin{figure*}
    \includegraphics[width=0.7\linewidth]{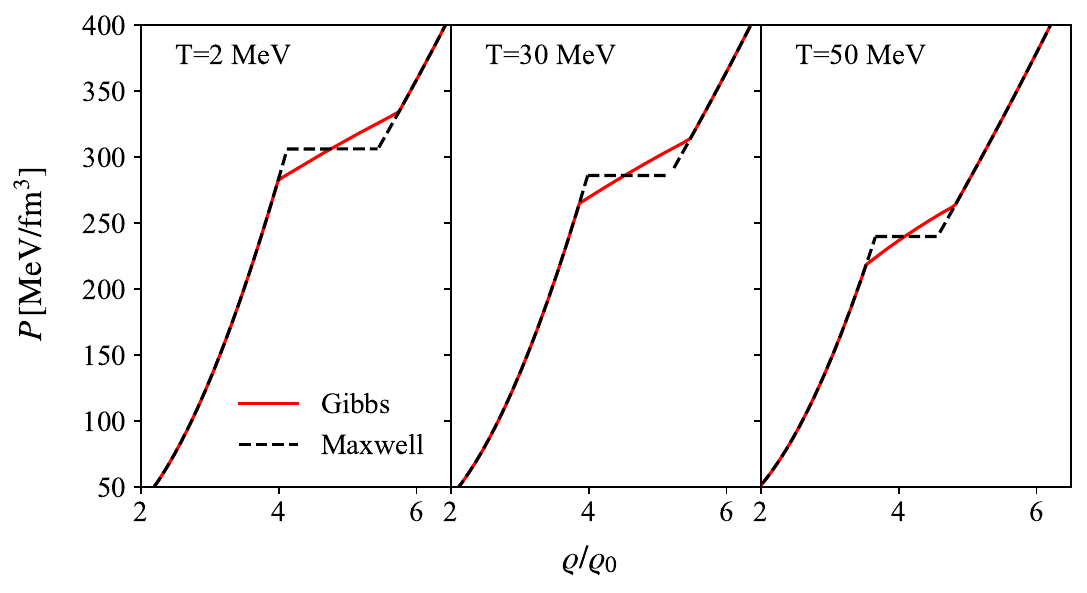}
    \caption{
    Comparison of the equations of state obtained using global
    lepton-number conservation (Gibbs construction; solid curves)
    and a local fixed-lepton-fraction prescription (Maxwell
    construction; dashed curves) at three representative
    temperatures.  Differences are concentrated in the
    hadron--quark coexistence region.
    }
    \label{fig:Max_Gibbs}
  \end{figure*}

Finally, we assess the sensitivity of the results to the treatment of
lepton-number conservation during the first-order transition.
Figure~\ref{fig:Max_Gibbs} compares the equations of state obtained
with the generalized Gibbs construction, in which electron-lepton
number is conserved globally, with those obtained using the
Maxwell-type prescription in which the prescribed lepton fraction is
imposed locally in each phase.

For an isothermal transition, the distinction is visible most clearly
inside the coexistence interval.  The Maxwell construction produces
the familiar constant-pressure coexistence behavior, whereas in the
Gibbs construction the pressure evolves through the mixed phase
because lepton number can be redistributed between its hadronic and
quark components.  Outside the coexistence region both constructions
reduce to the same underlying pure-phase equations of state, and the
differences consequently disappear.

For the parameter sets considered here, the quantitative differences
in $P(\varrho_B)$ are confined to a relatively narrow density
interval.  As a result, bulk stellar observables such as the
mass--radius relation are only weakly affected by replacing the Gibbs
construction by its Maxwell counterpart.  At first sight, this might
suggest that the choice of phase-equilibrium prescription is of
limited astrophysical importance.

\begin{figure*}
    \includegraphics[width=0.45\linewidth]{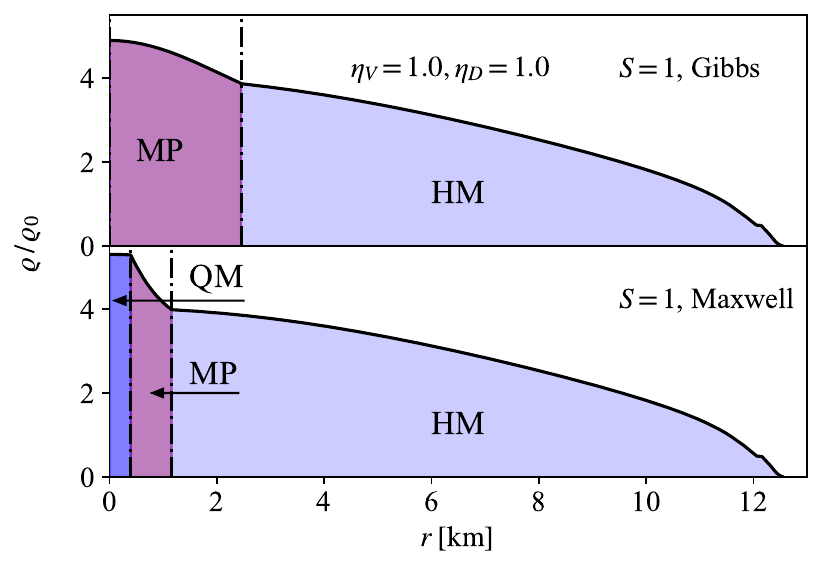}
    \includegraphics[width=0.45\linewidth]{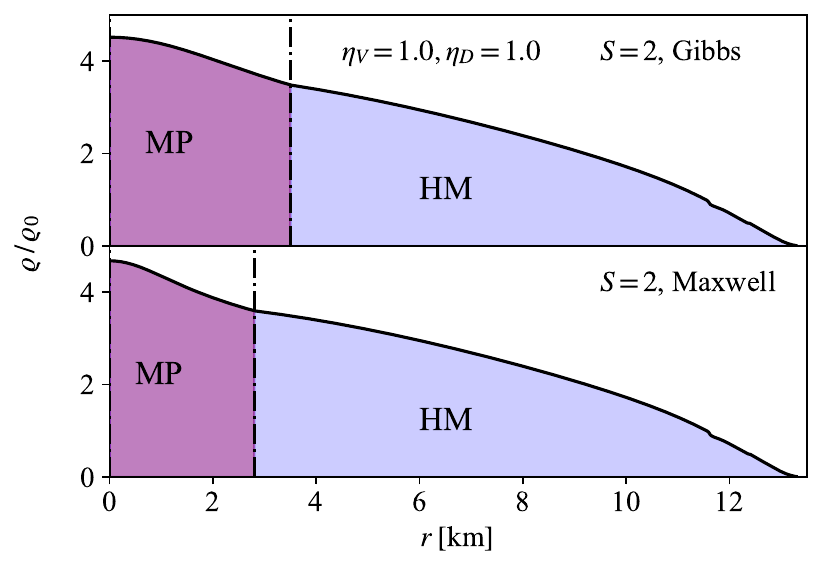}
    \caption{
    Internal profiles of maximum-mass isentropic hybrid-star
    configurations with trapped electron neutrinos
    ($Y_{Le}=0.4$).  The left and right panels correspond to
    $S=1$ and $S=2$, respectively.  The upper panels show the
    Gibbs construction with globally conserved electron-lepton
    number, while the lower panels show the Maxwell construction
    with a locally fixed lepton fraction. Shading indicates pure quark, mixed, and hadronic regions.  Although the global
    stellar properties are similar, the Gibbs construction produces
    a substantially more extended mixed-phase region.
    }
\label{fig:profile}
\end{figure*}

  The internal stellar structure, however, leads to a different
conclusion.  Figure~\ref{fig:profile} shows the radial density and
phase profiles of maximum-mass configurations at $S=1$ and $S=2$
for the two phase constructions.  Although their global stellar
properties are similar, the spatial distribution of the phases is
not.  In particular, the mixed-phase region is substantially more
extended when electron-lepton number is conserved globally.

This difference follows directly from the additional freedom
available in the Gibbs construction.  The local lepton fractions of
the hadronic and quark components can change independently while
their weighted average maintains the prescribed total $Y_{Le}$.
The system can therefore remain on the coexistence surface over a
larger range of densities.  In the Maxwell construction, by contrast,
the local lepton constraint reduces this freedom and restricts the
range over which the two phases can coexist along the isentropic
trajectory.

This observation illustrates an important limitation of using global
stellar observables alone to assess uncertainties in the treatment of
the phase transition.  {\it Two constructions that generate nearly
  indistinguishable mass--radius sequences can nevertheless predict
  rather different core compositions}.  The distinction can become
important for observables governed by local microscopic processes.
Neutrino opacities~\cite{Hofmann2025}, weak equilibration rates, thermal
and electrical transport~\cite{Alford:2014doa}, bulk
viscosity~\cite{Hernandez:2024rxi,Alford:2024tyj,Alford:2025tbp,Hernandez:2025zxw},
and the dynamics of phase conversion~\cite{Mintz2010,Mintz2010b,Bombaci2009,Bombaci:2010PTPS}  all depend on the volume
occupied by the individual phases and by their coexistence region.

The comparison therefore separates two levels of robustness.  The
global hydrostatic properties of the configurations are relatively
insensitive to whether lepton number is treated globally or locally
during phase coexistence.  The internal phase structure is
substantially more sensitive.  This distinction is likely to become
particularly relevant when the present equations of state are used in
dynamical calculations of proto-neutron stars or binary neutron-star
merger remnants.

\subsection{Sensitivity to the diquark coupling}
\label{sec:GD_discussion}

The foregoing comparisons were performed for the baseline diquark
coupling $\eta_D=G_D/G_S=1$.  To test the dependence on the pairing
interaction, we now consider $\eta_D=1.25$, leaving the vacuum
parameters and the hadronic model unchanged.  The two settings should
be viewed as representative choices for the pairing strength rather
than as an uncertainty interval derived from QCD.  An increase in
$G_D$ generally enhances the 2SC gap and changes the condensation
energy, thereby modifying the competition between the hadronic and
quark phases.  It can also change the amount of matter in the mixed
phase and the size of the pure quark core.

\begin{figure*}[tp]
\centering
\includegraphics[width=.8\textwidth]{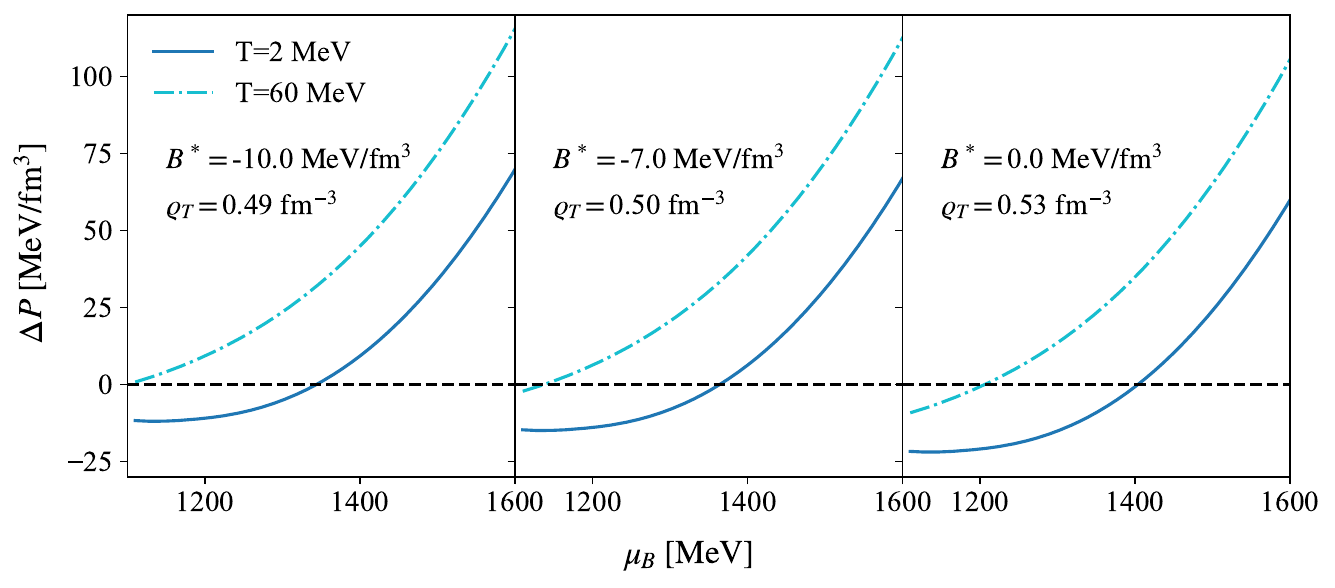}
\caption{Pressure difference $\Delta P=P_Q-P_H$ versus baryon
chemical potential for $\eta_V=1$ and stronger 2SC pairing,
$\eta_D=1.25$, in neutrino-transparent matter.  Three values of $B^*$ are compared; the solid
and dashed curves correspond to $T=2$ and $60$~MeV, respectively.
The different offsets move the crossings with $\Delta P=0$,
which mark mechanical coexistence.  Compare Fig.~\ref{fig:Bag}
for $\eta_D=1$.}\label{fig:Bag2}
\end{figure*}

Figure~\ref{fig:Bag2} displays the pressure competition in the
stronger-pairing case.  As in Fig.~\ref{fig:Bag}, we use
$\Delta P=P_Q-P_H$, so that a zero denotes a candidate coexistence
point and positive values favor quark matter.  In both figures the
quark pressure is shifted by $-B^*$; thus, a negative $B^*$ raises the
quark pressure.  The values of $B^*$ displayed in the two figures are
not the same.  The value of $B^*$ consistent with our prescription for $\eta_D=1.25$ is $B^*=-7\,\rm MeV/fm^3$, associated with a transition density at low temperature of $\varrho_{\rm tr}=0.50\,\rm fm^{-3}$. Increasing the diquark coupling allows for relatively lower onsets of deconfinement. The plots
therefore demonstrate the allowed adjustment of the transition
position within each pairing setup; their crossing points should not
be interpreted as an isolated effect of changing $G_D$ at a fixed
deconfinement offset.

\begin{figure}[tp]
    \centering
    \includegraphics[width=.9\linewidth]{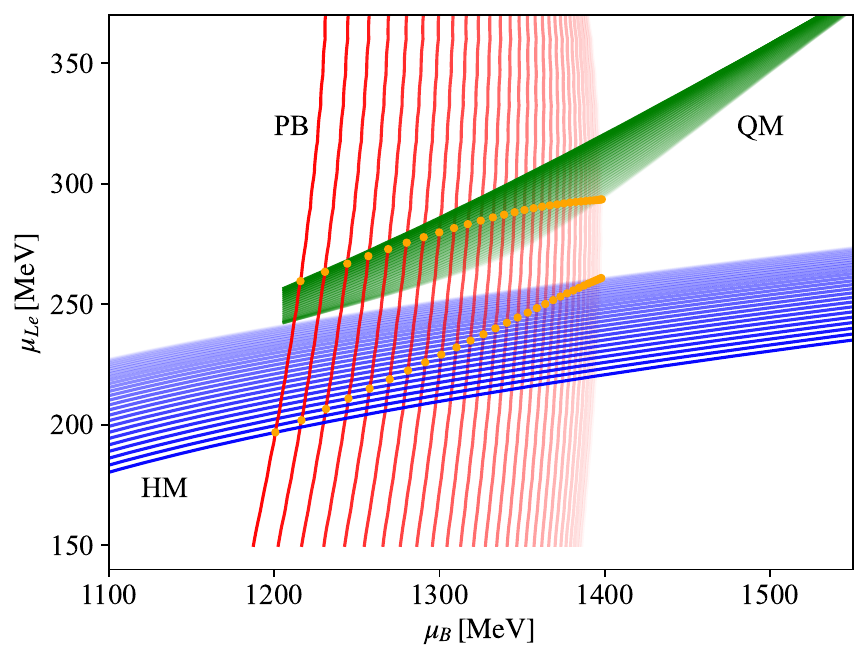}
    \caption{Phase-coexistence construction in the
    $(\mu_B,\mu_{Le})$ plane for $Y_{Le}=0.4$, $\eta_V=1$, and
    $\eta_D=1.25$.  The coexistence curve separates the regions
    of stable hadronic and quark matter; the lepton-fraction
    trajectories determine the mixed-phase endpoints.  Compare
    Fig.~\ref{fig:chempot_proj}.}
    \label{fig:chempot_proj2}
\end{figure}

The phase-equilibrium construction for $\eta_D=1.25$ is shown in
Fig.~\ref{fig:chempot_proj2}.  The phase boundaries again occupy a
curve in the $(\mu_B,\mu_{Le})$ plane, rather than a single value of
$\mu_B$.  A prescribed global $Y_{Le}$ then selects the entrance to
and exit from a coexistence interval.  Hence the two-charge Gibbs
construction continues to apply in the stronger-pairing case; changing
$G_D$ alters the location of the coexistence region but does not
remove its additional degree of freedom.

\begin{figure}[tp]
    \centering
    \includegraphics[width=\linewidth]{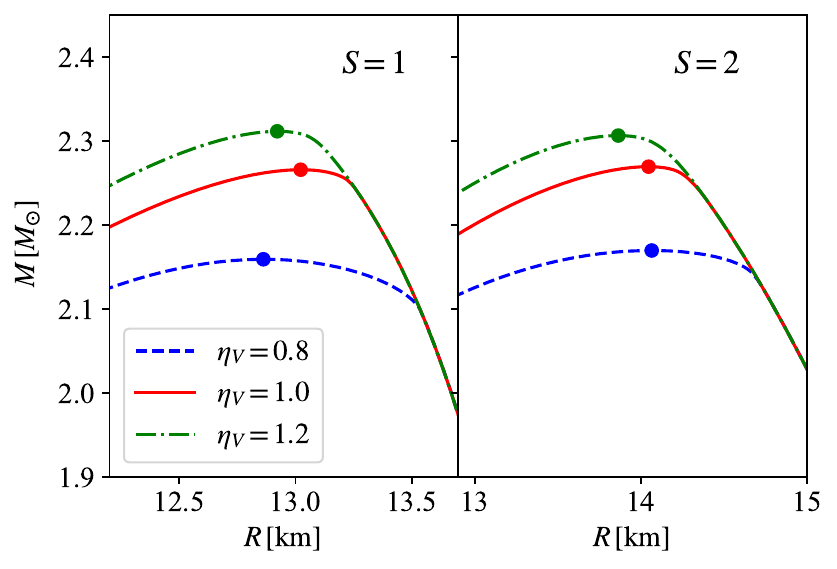}
    \caption{Mass--radius sequences for lepton-rich isentropic
    hybrid stars with $Y_{Le}=0.4$, $\eta_D=1.25$, and
    $\eta_V=0.8,1.0,1.2$.  The left and right panels show
    $S=1$ and $S=2$, respectively.  Dots identify the maximum
    masses along the displayed sequences.}
    \label{fig:MR_gv_comp2}
\end{figure}

We next assess whether the response to the repulsive vector interaction
is altered by stronger pairing.  Figure~\ref{fig:MR_gv_comp2}
compares the isentropic mass--radius sequences for the same vector
couplings considered in Sec.~\ref{sec:results_vector}.  In both the
$S=1$ and $S=2$ sequences, larger $\eta_V$ supports a larger maximum
mass.  Thus, the expected stiffening by the vector interaction remains
visible when the diquark coupling is increased.  The pairing and
vector couplings, however, affect the equation of state in different
ways: the former changes the 2SC condensate and phase equilibrium,
whereas the latter primarily changes the high-density repulsion.

For a quantitative comparison across the quark-model variants, we
introduce the fractional change in the maximum gravitational mass,
\begin{equation}
  \delta_M(S) \equiv
  \frac{M_{\max}^{(S)}-M_{\max,\mathrm{ref}}^{(S)}}
       {M_{\max,\mathrm{ref}}^{(S)}} .
  \label{eq:mass_fractional_change}
\end{equation}
The reference configuration is the sharp-cutoff NJL model with
$\eta_D=\eta_V=1$ and DDLS50 hadronic matter, as in
Ref.~\cite{Sabatucci2026}; its maximum masses are
$M_{\max,\mathrm{ref}}^{(S=1)}=2.36\,M_\odot$ and
$M_{\max,\mathrm{ref}}^{(S=2)}=2.37\,M_\odot$.
Table~\ref{tab:masses} summarizes $100\delta_M$ for lepton-rich
matter at fixed $Y_{Le}=0.4$.  For the baseline pairing strength,
changing $\eta_V$ from $0.8$ to $1.2$ changes the maximum mass by
about $-1.3\%$ to $+1.3\%$ relative to the reference model.
Stronger pairing and the RG-consistent regularization produce larger
shifts for certain parameter combinations.  The underlying maximum
masses, radii, and mixed-phase boundaries are tabulated in
Appendix~\ref{chapter:addition}.  As the available maximum masses
are tabulated to $0.01\,M_\odot$, the percentages should be
interpreted as approximate.

\begin{table}[tp]
  \centering
  \caption{Percentage change $100\delta_M(S)$ of the maximum
  gravitational mass relative to the DDLS50--NJL sharp-cutoff
  reference model with $\eta_D=\eta_V=1$ at the same entropy.
  The matter is neutrino trapped, with $Y_{Le}=0.4$.  Values are
  calculated from the rounded masses in Table~\ref{tab:big_table}.}
  \label{tab:masses}
  \vspace{0.5cm}
  \begin{tabular}{lccrr}
    \hline\hline
    NJL scheme & $\eta_D$ & $\eta_V$ & $S=1$ [\%] & $S=2$ [\%] \\
    \hline
    Cutoff & 1.0  & 0.8 & $-1.3$ & $-1.3$ \\
           & 1.0  & 1.0 & $\phantom{-}0.0$ & $\phantom{-}0.0$ \\
           & 1.0  & 1.2 & $+1.3$ & $+1.3$ \\
           & 1.25 & 0.8 & $-8.5$ & $-8.4$ \\
           & 1.25 & 1.0 & $-3.8$ & $-4.2$ \\
           & 1.25 & 1.2 & $-2.1$ & $-3.0$ \\
    \hline
    RG     & 1.0  & 1.0 & $-1.7$ & $-1.7$ \\
           & 1.25 & 1.0 & $-8.5$ & $-8.4$ \\
    \hline\hline
  \end{tabular}
\end{table}

\begin{figure}[tp]
  \centering
  \includegraphics[width=\linewidth]{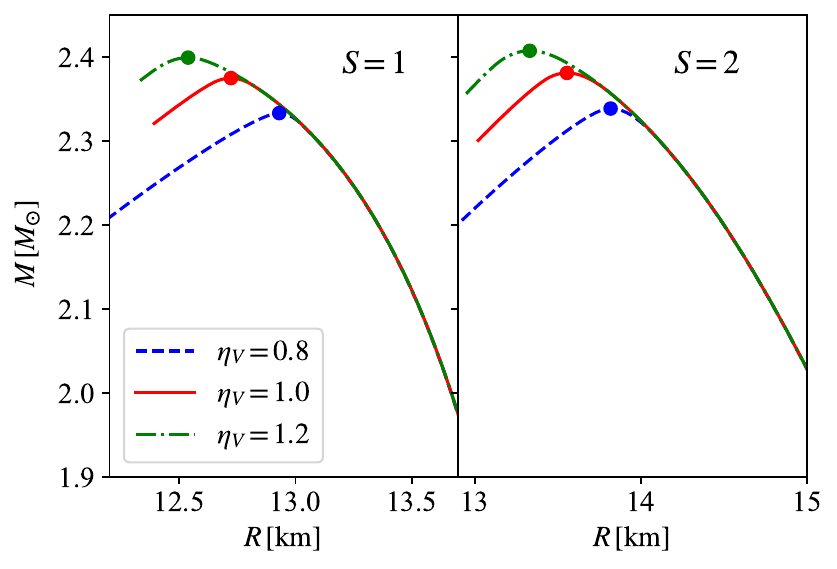}
  \par\vspace{0.7em}
  \includegraphics[width=\linewidth]{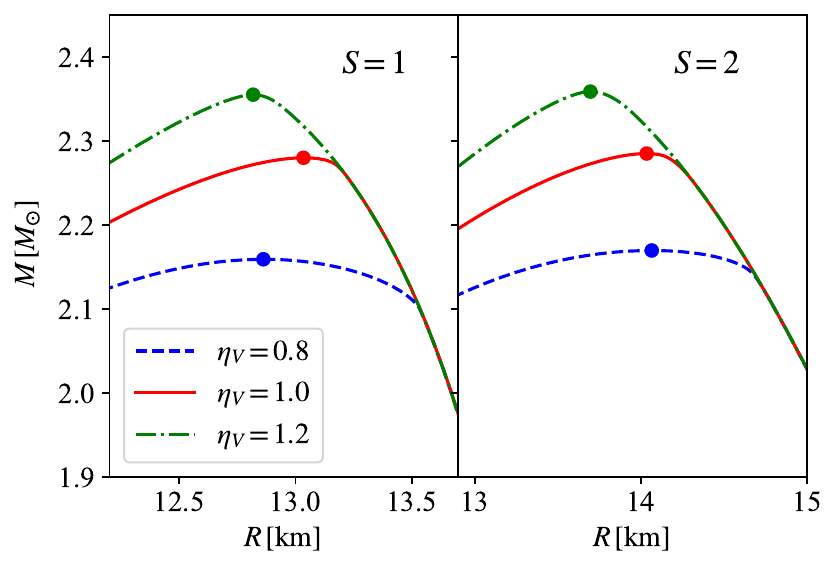}
  \caption{Mass--radius sequences at fixed phenomenological pressure
  shift $B^*$ for $\eta_V=0.8,1.0,1.2$, with trapped electron
  neutrinos ($Y_{Le}=0.4$).  The upper row uses $\eta_D=1$ and
  $B^*=-15\,\mathrm{MeV\,fm^{-3}}$; the lower row uses
  $\eta_D=1.25$ and $B^*=-5\,\mathrm{MeV\,fm^{-3}}$.
  The left and right panels in each row correspond to $S=1$ and
  $S=2$.  Compare with Figs.~\ref{fig:MR_gv_comp} and
  \ref{fig:MR_gv_comp2}, where $B^*$ is readjusted separately
  for each $\eta_V$ according to the prescription in
  Sec.~\ref{sec:results_vector}.  In both comparisons, larger
  vector repulsion supports larger maximum masses.}
  \label{fig:MR_gv_comp_fixedB}
\end{figure}

Because our default prescription readjusts $B^*$ for each quark
coupling, a change of $\eta_V$ is accompanied by a change in the
phenomenological pressure shift.  We therefore carry out an
additional control comparison holding $B^*$ fixed across the three
vector couplings at each value of $\eta_D$.
Figure~\ref{fig:MR_gv_comp_fixedB} shows these mass--radius
sequences.  The increasing order of the maximum masses with
$\eta_V$ is preserved at both pairing strengths, indicating that
this trend is not generated solely by the adjustment of $B^*$.
The detailed locations and heights of the maxima do change when
$B^*$ is retuned; thus, the fixed-$B^*$ experiment provides a
robustness check on the trend, not a determination of the
absolute mass independent of the deconfinement prescription.

\begin{figure*}[tp]
    \centering
    \includegraphics[width=.7\linewidth]{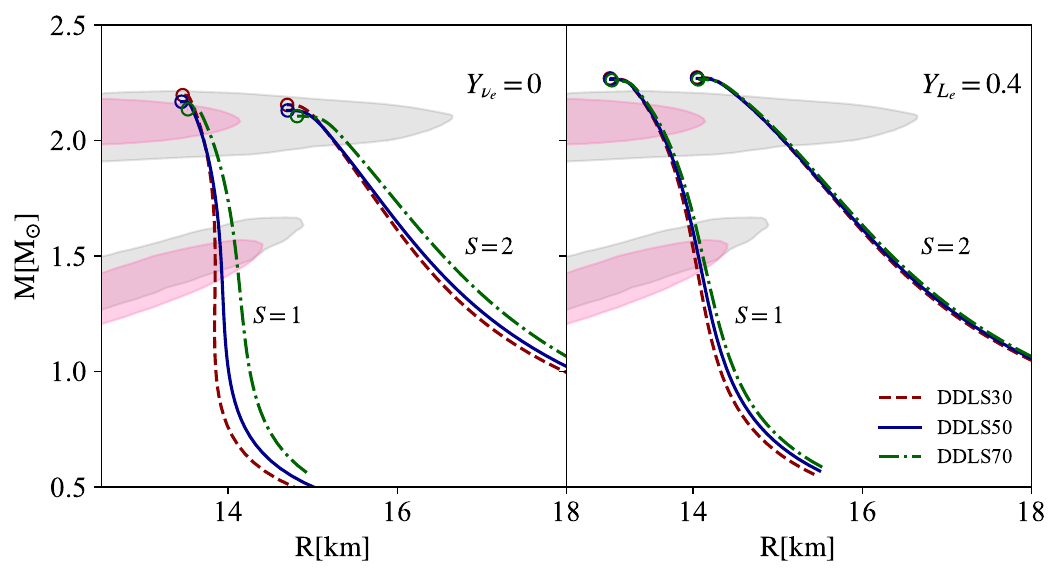}
    \caption{Mass--radius relations for the three DDLS symmetry-energy
    slopes $L_{\rm sym}=30,50,70$~MeV at $\eta_V=1$ and
    $\eta_D=1.25$.  Neutrino-transparent matter is shown on the
    left and lepton-rich matter with $Y_{Le}=0.4$ on the right;
    both $S=1$ and $S=2$ sequences are included.  The reduced
    spread in the trapped case persists for stronger 2SC pairing.}
    \label{fig:MR_DDLS_comp2}
\end{figure*}

A complementary hadronic uncertainty is shown in
Fig.~\ref{fig:MR_DDLS_comp2}, where the symmetry-energy slope is
varied at $\eta_D=1.25$.  As in the baseline pairing case, the
neutrino-trapped configurations show a smaller spread with
$L_{\rm sym}$ than the corresponding neutrino-transparent sequences.
This supports the interpretation of Sec.~\ref{sec:results_Lsym}:
the reduction of isospin asymmetry in lepton-rich hadronic matter
mitigates the dependence on the nuclear symmetry-energy slope, even
when the quark pairing interaction is changed.

\begin{figure*}[tp]
    \centering
   \includegraphics[width=.45\linewidth]{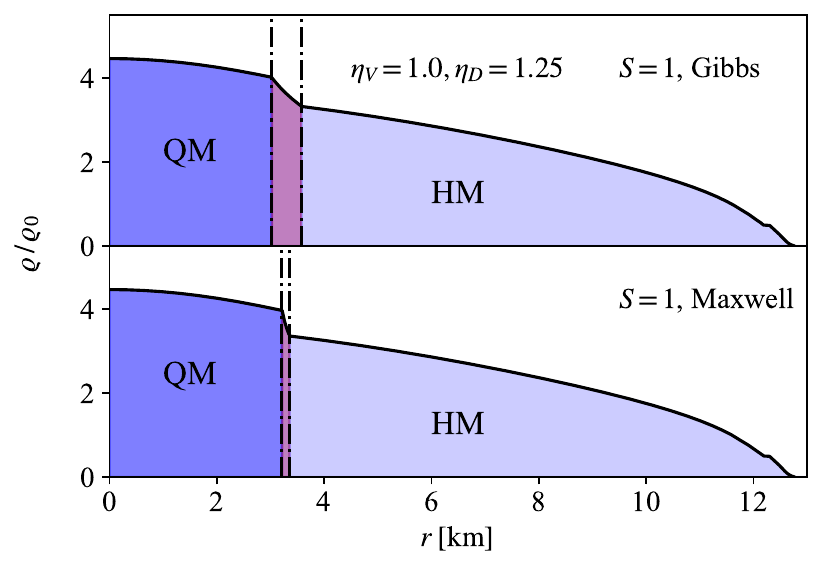}
\includegraphics[width=.45\linewidth]{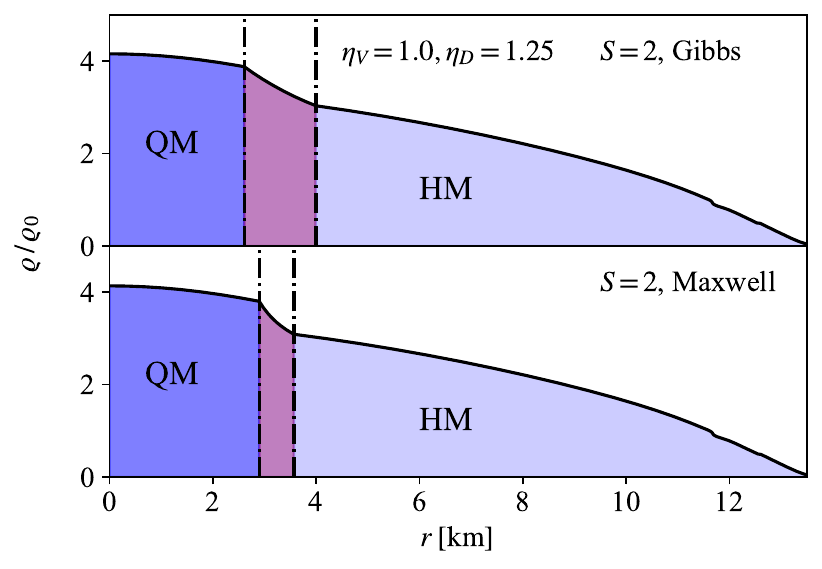}
    \caption{Radial profiles of maximum-mass configurations with
    $Y_{Le}=0.4$, $\eta_V=1$, and $\eta_D=1.25$.
    The upper and lower panels within each plot use the Gibbs and
    Maxwell lepton-number constraints, respectively.  The shaded
    intervals distinguish quark matter, mixed matter, and hadronic
    matter.  Compare Fig.~\ref{fig:profile} for $\eta_D=1$.}
\label{fig:profile2}
\end{figure*}

The internal profiles make the influence of the pairing interaction
more evident.  Figure~\ref{fig:profile2} shows maximum-mass stars for
$S=1$ and $S=2$ at $\eta_D=1.25$, with both the global-lepton
(Gibbs) and local-lepton (Maxwell) constructions.  The radial extent
of the quark, mixed, and hadronic regions changes between the two
prescriptions, while the extent of the pure quark core also depends
on the pairing setup (compare Fig.~\ref{fig:profile}).  These
comparisons reinforce a recurring point: a modification of the pairing
sector may be more apparent in the phase composition than in the
global mass--radius curve.  Because the pressure offset can also move
the deconfinement boundary, the separate effects of pairing and
$B^*$ should be distinguished when comparing models quantitatively.

\subsection{Effect of RG-consistent NJL regularization}
\label{sec:RG_comparison}

Conventional NJL calculations regularize the quark momentum integrals
using a cutoff fitted to vacuum observables.  When the chemical
potential or temperature becomes a substantial fraction of this
cutoff, truncating the medium contributions at the same scale can
produce artifacts, particularly in color-superconducting gaps.
Renormalization-group (RG)-consistent treatments provide a way of
separating the vacuum cutoff from the scale at which the medium
contributions are integrated and of removing the additional medium
divergences associated with pairing~\cite{Gholami_RGfirst,Gholami2025}.
We use the resulting quark equation of state to reassess neutrino-rich
hybrid stars for $\eta_D=1$ and $1.25$, with $\eta_V=1$.

We briefly summarize the prescription to make the comparison
reproducible. Further details can be found in Ref.~\cite{Gholami_RGfirst, Gholami2025} and in the open-source NJL module~\cite{hofmann_2026_18249033}, which was released as part of the latest version of \textit{Calliope}, the MUSES Calculation Engine~\cite{Jahan:2026hvs}.

Denote the original vacuum cutoff by
$\Lambda_0=602.3$~MeV (the $\Lambda$ of
Eq.~\eqref{eq:Pressure}), and introduce a higher auxiliary scale
$\Lambda_{\rm UV}>\Lambda_0$.  Let $\boldsymbol\mu$ collect the
chemical potentials and $\boldsymbol\chi$ the condensates, and
write the unregularized mean-field integrand as
$\mathcal A(p;\boldsymbol\mu,T,\boldsymbol\chi)$.  Its vacuum
part is
$\mathcal A_{\rm vac}(p;\boldsymbol\chi)
 =\mathcal A(p;\boldsymbol\mu=0,T=0,\boldsymbol\chi)$.
The effective grand potential can then be expressed as
\begin{eqnarray}
\Omega_{\rm eff}^{\rm R}
&=&\int_0^{\Lambda_0}\frac{p^2dp}{2\pi^2}
    \mathcal A_{\rm vac}(p;\boldsymbol\chi)\nonumber\\
& +&\int_0^{\Lambda_{\rm UV}}\frac{p^2dp}{2\pi^2}
 \left[\mathcal A(p;\boldsymbol\mu,T,\boldsymbol\chi)
       -\mathcal A_{\rm vac}(p;\boldsymbol\chi)\right]\nonumber\\
& + &\mathcal V(\boldsymbol\chi)+\delta\Omega .
\label{eq:RGpotential}
\end{eqnarray}
Here $\mathcal V$ contains the mean-field condensate terms.
Unlike in unpaired matter, the medium part in a paired state still
contains terms that diverge logarithmically with the auxiliary scale.
The counterterm $\delta\Omega$ removes those terms while preserving
the original vacuum parametrization.

For the minimal subtraction scheme of
Ref.~\cite{Gholami_RGfirst}, the counterterm can be written in the
present gap notation as
\begin{eqnarray}
\delta\Omega&=&-\frac{1}{\pi^2}
\ln\!\left(\frac{\Lambda_{\rm UV}}{\Lambda_0}\right)
\big[(\mu^2_{ur,dg}+\mu^2_{ug,dr})|\Delta_3|^2\nonumber\\
& + &(\mu^2_{ur,sb}+\mu^2_{ub,sr})|\Delta_2|^2
 + (\mu^2_{dg,sb}+\mu^2_{db,sg})|\Delta_1|^2\big],\nonumber\\
\label{eq:RGcounterterm}
\end{eqnarray}
with
\begin{equation}
\mu_{\alpha a,\beta b}=\tfrac12(\mu_{\alpha a}+\mu_{\beta b}).
\end{equation}
The minimal scheme differs from alternative RG-consistent
subtraction prescriptions by finite terms; consequently, the scheme
is part of the model specification.  In the 2SC phase only the
$\Delta_3$ condensate is nonzero.

An equivalent form explicitly displays the conventional contribution
and the additional high-momentum terms:
\begin{eqnarray}
\Omega_{\rm eff}^{\rm R}
&=&\int_0^{\Lambda_0}\frac{p^2dp}{2\pi^2}
   \mathcal A(p;\boldsymbol\mu,T,\boldsymbol\chi)
  +\mathcal V(\boldsymbol\chi)\nonumber\\
 &+&\int_{\Lambda_0}^{\Lambda_{\rm UV}}
 \frac{p^2dp}{2\pi^2}
 \left[\mathcal A(p;\boldsymbol\mu,T,\boldsymbol\chi)
 -\mathcal A_{\rm vac}(p;\boldsymbol\chi)\right]
 +\delta\Omega .\nonumber\\
\label{eq:RGdecomposition}
\end{eqnarray}
The numerical comparison presented here uses
$\Lambda_{\rm UV}=5\Lambda_0$.  For each regularization the
gaps, mean fields, and neutrality chemical potentials are solved
self-consistently before phase matching.  The hadronic input and the
thermodynamic lepton constraints remain as described in the previous
sections.  The effective deconfinement shift $B^*$ remains a separate,
phenomenological input: a change of regularization does not by itself
fix the hadron--quark pressure normalization.

\begin{figure}[tp]
    \centering
    \includegraphics[width=\linewidth]{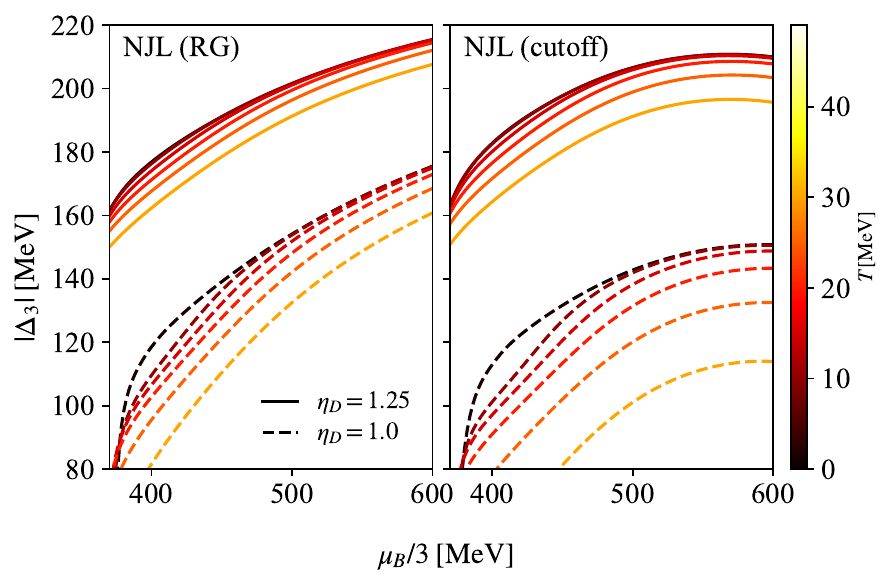}
    \caption{Magnitude of the 2SC pairing gap as a function of quark
    chemical potential for different temperatures (indicated by the
    color scale).  Solid curves represent $\eta_D=1.25$ and dashed
    curves $\eta_D=1$.  The left panel uses RG-consistent
    regularization and the right panel the conventional momentum
    cutoff.  The RG-consistent treatment reduces the spurious
    high-chemical-potential suppression of pairing.}
    \label{fig:gapRG}
\end{figure}

Figure~\ref{fig:gapRG} compares the 2SC gap as a function of quark
chemical potential and temperature.  In the conventional cutoff
scheme the gap tends to turn down  or decrease toward larger chemical
potentials, whereas the RG-consistent calculation mitigates this
cutoff-induced suppression.  The distinction is evident for both
$\eta_D=1$ and $1.25$, and the larger diquark coupling gives the
larger gaps.  Since the condensation energy affects phase equilibrium,
these differences can propagate to the hybrid equation of state.

\begin{figure*}[tp]
    \centering
    \includegraphics[width=.8\textwidth]{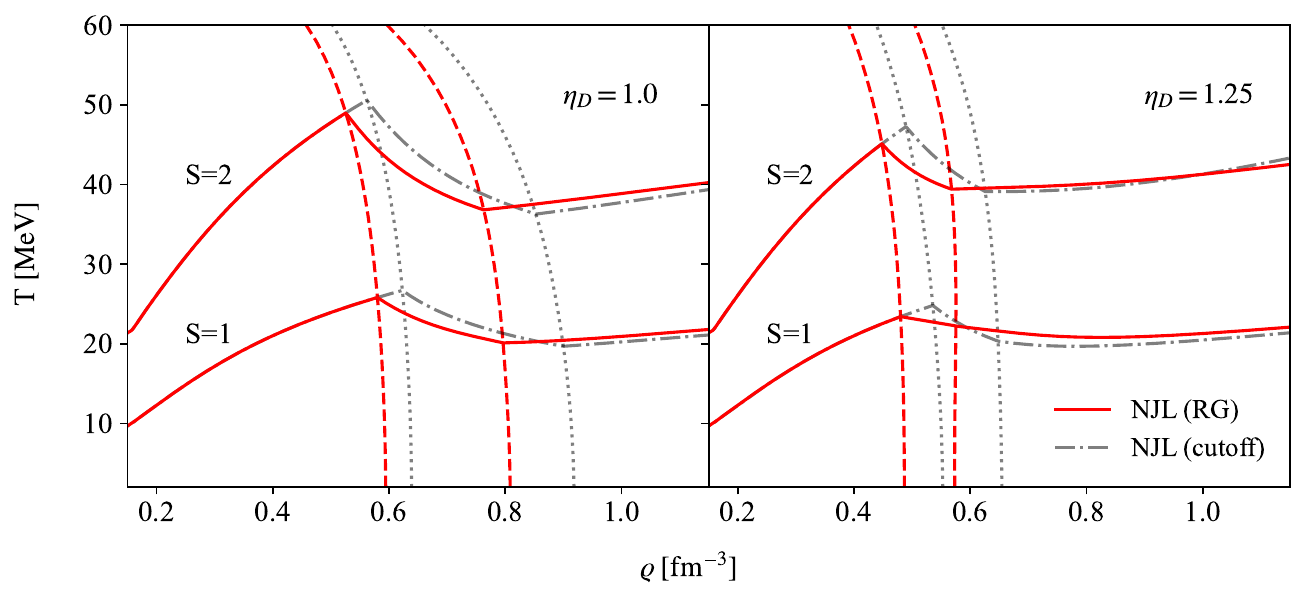}
    \caption{Temperature versus baryon density for lepton-rich
    ($Y_{Le}=0.4$) isentropic matter, with $S=1$ and $2$.
    The left and right panels correspond to $\eta_D=1$ and $1.25$;
    $\eta_V=1$ throughout.  Solid red curves use RG-consistent
    regularization, while dot-dashed gray curves use the conventional
    cutoff.  Vertical/steep segments indicate phase-transition
    boundaries along the plotted trajectories.}
    \label{fig:PD_RGcomp}
\end{figure*}

The consequences for lepton-rich matter are summarized in
Fig.~\ref{fig:PD_RGcomp}.  At fixed $Y_{Le}=0.4$ the different
regularizations modify both the density interval crossed by an
isentrope in the phase-coexistence region and the temperature profile
along that isentrope.  The RG-consistent curves show a transition at
lower densities in the displayed comparisons, and the changes are
visible for both diquark couplings.  The locations of the boundaries
should, however, be interpreted together with the effective pressure
matching prescription: the regularization and $B^*$ are logically
independent inputs to the hybrid construction. The values of $B^*$ in the RG-consistent NJL, for both choices of $\eta_D$ are the same as their counterparts in the model without RG corrections. However they are associated to somehow lower onset densities, $\varrho_{\rm tr}=0.54\,\rm fm^{-3}$ and $\varrho_{\rm tr}=0.47\,\rm fm^{-3}$ for $\eta_D=1$ and $\eta_D=1.25$ respectively.
 
\begin{figure}[tp]
    \centering
    \includegraphics[width=\linewidth]{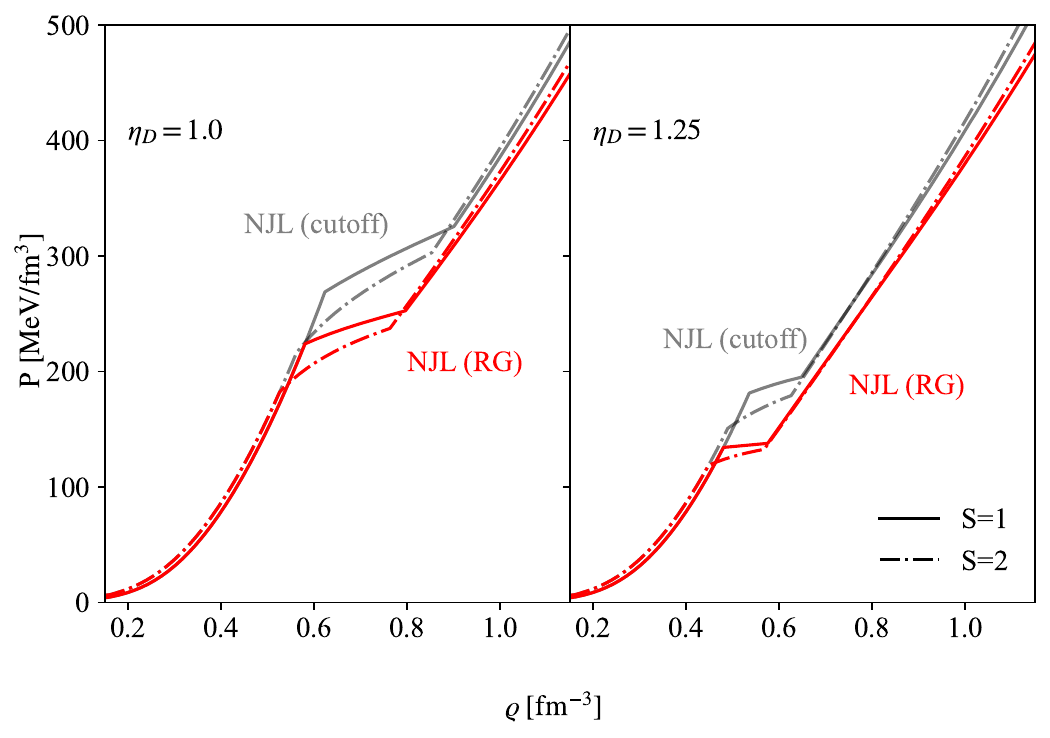}
    \caption{Isentropic pressure--density relations for trapped
    electron neutrinos ($Y_{Le}=0.4$) at $\eta_V=1$.
    The left and right panels correspond to $\eta_D=1$ and $1.25$,
    with $S=1$ (solid curves) and $S=2$ (dashed curves).
    Red curves use RG-consistent regularization; gray curves
    use the conventional cutoff.}
    \label{fig:EOS_RGcomp}
\end{figure}

Figure~\ref{fig:EOS_RGcomp} exhibits the corresponding pressure--density
relations.  The two regularization schemes agree more closely in the
hadronic regime, whose microscopic equation of state is unchanged,
than in the density interval containing the transition and the quark
phase.  The comparison shows that the regularization prescription is
an additional source of microscopic uncertainty, distinct from
varying the vector or diquark coupling at fixed scheme.  In particular,
a larger pairing gap should not be equated automatically with a
uniformly stiffer hybrid equation of state, because the transition
position and coexistence interval also enter.

\begin{figure*}[tp]
    \centering
    \includegraphics[width=0.8\linewidth]{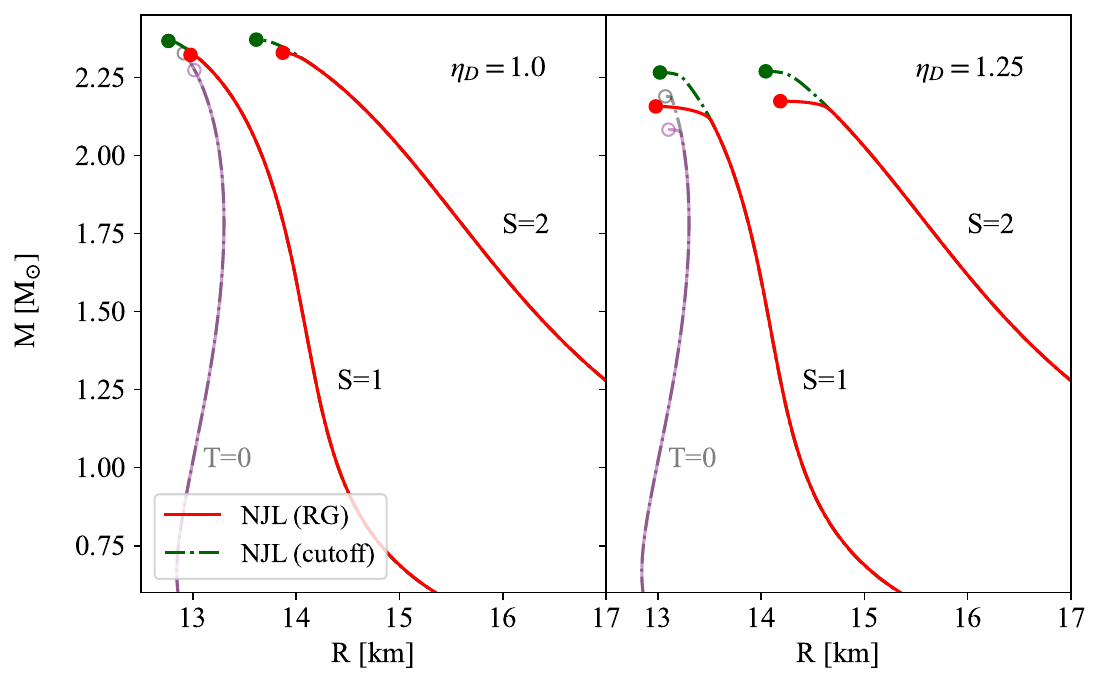}
    \caption{Mass--radius sequences for trapped-neutrino isentropes
    ($Y_{Le}=0.4$) with $S=1$ and $S=2$ for $\eta_D=1$ (left) and
    $\eta_D=1.25$ (right), at $\eta_V=1$.  RG-consistent and
    conventional-cutoff results are shown in red and gray,
    respectively.  The thin cold ($T=0$) sequences included in the
    figure provide a reference for the finite-entropy effects.}
    \label{fig:MR_RGcomp}
\end{figure*}

The resulting stellar sequences are compared in
Fig.~\ref{fig:MR_RGcomp}.  The mass--radius curves remain relatively
close for the parameter choices shown, especially when compared with
the more direct changes of the transition boundary and gap functions.
The same figure illustrates the distinct radii of the cold and hot
isentropic sequences.  This provides another example in which
macroscopic hydrostatic observables can show only moderate differences
between microphysical prescriptions that yield different internal
compositions.

\begin{figure*}[tp]
    \centering
    \includegraphics[width=.45\linewidth]{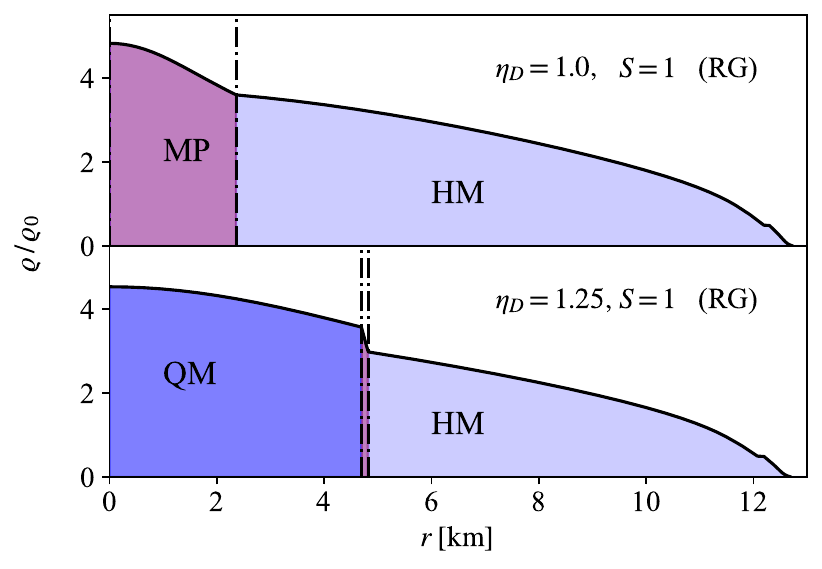}
    \includegraphics[width=.45\linewidth]{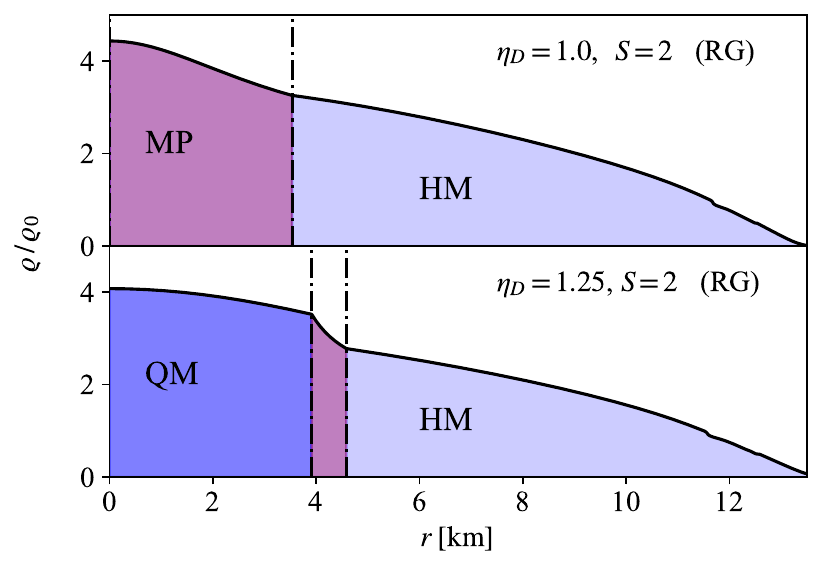}
    \caption{Radial matter profiles of maximum-mass hybrid stars
    constructed with the RG-consistent NJL model, trapped electron
    neutrinos ($Y_{Le}=0.4$), and $\eta_V=1$.  The upper and
    lower panels of each plot show $\eta_D=1$ and $\eta_D=1.25$,
    respectively.  Shading indicates pure quark, mixed, and hadronic
    regions.  Both $S=1$ and $S=2$ are illustrated.}
    \label{fig:profile_RG}
\end{figure*}

Finally, Fig.~\ref{fig:profile_RG} compares the radial phase profiles
of maximum-mass configurations calculated in the RG-consistent scheme.
For each entropy, the upper and lower panels show $\eta_D=1$ and
$1.25$, respectively.  Stronger pairing changes the radial boundaries
of the quark and mixed regions, with a more extended quark-containing
core in the displayed configurations.  This figure compares two
pairing strengths within the RG-consistent scheme; it should not be
confused with a direct profile-by-profile comparison of regularization
schemes.

The RG-consistent extension strengthens a key conclusion of this
study.  The high-density pairing sector and the pressure matching
can appreciably influence the transition boundary and phase
composition.  Yet the induced changes in the static stellar sequences
need not be comparably prominent.  Predictions for local transport
and phase-conversion phenomena therefore require tighter control of
the regularization and transition prescription than may be suggested
by bulk mass--radius curves alone.

\section{Summary and Conclusions}
\label{sec:discussion}
In this work we have extended the analysis of
Ref.~\cite{Sabatucci2026} to examine the robustness of the
hadron--quark phase transition and of the resulting hot hybrid-star
configurations against several sources of model dependence.  The
previous study established the thermodynamics of isentropic hybrid
matter with 2SC superconducting quark matter and showed
that trapped electron neutrinos delay the onset of deconfinement and
modify the properties of hot stellar configurations.  Here we have
used that calculation as a reference and varied, in turn, the
high-density quark interaction, the isovector properties of the
hadronic equation of state, the diquark pairing strength, the regularization of the NJL model,
and the treatment of lepton-number conservation across the
first-order transition.

On the quark-matter side, we considered variations of the repulsive
vector coupling $\eta_V=G_V/G_S$.  As expected, increasing
$\eta_V$ stiffens the quark-matter equation of state and leads to
larger maximum masses of hybrid configurations.  Within the interval
$0.8\leq\eta_V\leq1.2$ considered here, however, the corresponding
changes of the maximum mass remain moderate and do not alter the
qualitative effect of neutrino trapping.  In particular, the shift
of deconfinement toward higher density in lepton-rich matter persists
throughout the range of vector couplings studied.  The absolute
location of the transition and the high-density stiffness therefore
retain a quantitative quark-model dependence, whereas the direction
of the trapping effect is considerably more robust.

We have also clarified the role of the effective pressure shift
$B^*$ entering the NJL equation of state.  Since the NJL model does
not incorporate confinement, it does not determine from first
principles the relative normalization of the hadronic and quark
pressures at deconfinement.  The parameter $B^*$ is therefore used
here as an effective prescription for placing the deconfinement
transition rather than as a microscopic interaction parameter.
The pressure difference $P_Q-P_H$ provides a direct representation
of this role: hadronic matter is favored while $P_H>P_Q$, the
coexistence condition is reached at equal pressure, and quark matter
becomes favored once $P_Q>P_H$.  The variation of $B^*$ consequently
quantifies an uncertainty associated with the effective modeling of
confinement physics that lies outside the NJL description itself.
Our additional comparison at fixed $B^*$ (Fig.~\ref{fig:MR_gv_comp_fixedB})
confirms that larger quark vector repulsion increases the maximum
stellar mass in both pairing setups even without readjusting the
phenomenological offset, although the precise peak masses and radii
remain dependent on this choice.

A second robustness test concerned the hadronic equation of state.
We repeated the calculation for three members of the DDLS family
with symmetry-energy slopes
$L_{\rm sym}=30$, $50$, and $70$ MeV.  In neutrino-transparent
matter these models lead to a visible spread in the stellar
mass--radius relations, reflecting the sensitivity of strongly
neutron-rich matter to the density dependence of the nuclear
symmetry energy.  This spread is reduced when electron neutrinos are
trapped.  The effect has a simple physical origin: fixing a sizable
electron-lepton fraction increases the electron and proton
abundances and drives the hadronic phase toward greater isospin
symmetry.  Since the symmetry-energy contribution scales
approximately with the square of the isospin asymmetry, the
sensitivity to $L_{\rm sym}$ is correspondingly reduced.  Thus,
within the class of hadronic models considered here, neutrino
trapping partially suppresses rather than enhances the uncertainty
associated with the isovector sector.

The third source of uncertainty is conceptually different.  It
concerns the thermodynamic treatment of electron-lepton number during
phase coexistence.  Our Gibbs construction assumes local electric
neutrality but global conservation of electron-lepton number, allowing
lepton number to be redistributed between the hadronic and quark
components of the mixed phase.  We compared this with a Maxwell-type
construction in which the prescribed lepton fraction is imposed
locally in each phase.  The resulting pressure--density relations
differ mainly within the coexistence region, and these differences
have only a modest influence on the global mass--radius relation.
Static stellar observables alone may therefore suggest that the
distinction between the two constructions is relatively unimportant.

The internal stellar profiles show, however, that this conclusion
would be incomplete.  For maximum-mass isentropic configurations the
Gibbs construction produces a substantially more extended
mixed-phase region than the Maxwell prescription.  Global
lepton-number conservation permits the two phases to carry different
local lepton fractions while maintaining the prescribed total
$Y_{Le}$, allowing coexistence over a wider range of stellar
conditions.  The locally constrained Maxwell construction removes
this additional degree of freedom and correspondingly restricts the
mixed region.  Thus, equations of state that give rather similar
bulk stellar properties can nevertheless predict significantly
different internal phase compositions.

This distinction is potentially important because many properties
of hot compact stars depend on the local composition rather than
only on the integrated equation of state.  Weak equilibration rates,
neutrino opacities, thermal transport, viscosities, and the dynamics
of phase conversion can all depend sensitively on whether a given
radial interval consists of pure hadronic matter, pure quark matter,
or a coexistence region.  The relatively weak sensitivity of the
mass--radius relation to the phase construction should therefore not
be interpreted as implying an equally weak sensitivity of dynamical
or transport observables.

We further varied the diquark coupling from $\eta_D=1$ to
$\eta_D=1.25$.  Stronger pairing changes the 2SC condensate and the
phase-equilibrium construction and can noticeably modify the radial
extent of the quark-containing core.  The sensitivity to the quark
vector repulsion and the reduced dependence on $L_{\rm sym}$ in
neutrino-rich matter remain apparent in the stronger-pairing model.
In this comparison the effective deconfinement shift must be
specified alongside the pairing strength, because changes of $B^*$
can independently move the phase transition.

Finally, we examined the effect of replacing conventional NJL cutoff
regularization with an RG-consistent treatment.  Removing the
high-momentum cutoff artifacts mitigates the suppression of the 2SC
gap at large chemical potentials and changes the deconfinement
boundaries along isentropic trajectories.  For the displayed model
choices the resulting stellar mass--radius sequences are less
sensitive than the microscopic gaps and phase composition.  The
RG-consistent model is thus a useful extension of the microscopic
uncertainty analysis, but it does not eliminate the independent
phenomenological ambiguity represented by $B^*$, which has a different
physical origin. 

To conclude, our overall results reveal a useful hierarchy of uncertainties.
The qualitative tendency of trapped neutrinos to delay deconfinement
is robust against the variations of the quark vector interaction and
hadronic symmetry energy considered here.  The precise transition
density and the maximum mass remain quantitatively model dependent.
The largest qualitative uncertainty appears not necessarily in the
global stellar structure, but in the internal distribution of phases,
which is sensitive to how lepton-number conservation is implemented
during the transition.  {\it This separation between robust
  macroscopic trends and much more sensitive microscopic
  composition is one of the main conclusions of the present work.}

The calculations presented here still rely on several simplifying
assumptions.  Our equilibrium stellar models use spatially uniform
entropy per baryon and, in trapped matter, a prescribed electron-lepton
fraction, whereas realistic proto-neutron stars and merger remnants
have spatially and temporally varying entropy and lepton profiles.  We have restricted
the quark phase to 2SC matter and have neglected the muon-lepton
sector.  Our RG-consistent results employ the minimal subtraction
scheme and a finite auxiliary cutoff, so residual numerical and
scheme dependence should be assessed separately.  Furthermore, the location of deconfinement is modeled
phenomenologically through $B^*$ because confinement is absent from
the NJL framework.  The Gibbs and Maxwell constructions should
likewise be viewed as limiting descriptions of phase equilibrium; the
actual behavior may depend on surface tension, charge screening,
diffusion, and the timescale of phase conversion.

These limitations also indicate the natural next steps.  Of particular
interest is the incorporation of the present equations of state into
dynamical proto-neutron-star~\cite{Fischer:2017lag} and binary-merger
simulations~\cite{Most:2019,Bauswein:2019,Prakash:2021}, where the
different internal mixed-phase structures could affect neutrino
transport, dissipation, and post-merger evolution even when the
corresponding equilibrium mass--radius relations are nearly
indistinguishable.  Extending the microscopic description to
additional paired quark phases and treating the lepton sector and
phase conversion dynamically will allow the robustness identified here
at the level of equilibrium stellar structure to be tested in
transient compact-star environments.

\section*{Acknowledgment}

We thank M.~Alford, D.~Blaschke, H.~Gholami, A.~Harutyunyan and
M.~Oertel for discussions.  This work was supported by the
Polish National Science Centre (NCN) Grant 2023/51/B/ST9/02798.

\appendix
\section{Bag constant and transition density}\label{chapter:tables}
In the following, we report the values of $B^*$ and the corresponding transition densities for all the cases considered in this work.

Table~\ref{tab:1} lists the numerical values for the different hadronic EoSs, characterized by different values of the symmetry-energy slope parameter $L_{\rm sym}$, as well as different vector and diquark coupling strengths, $\eta_V=G_V/G_S$ and $\eta_D=G_D/G_S$. The table also includes results obtained using both regularization schemes of the NJL model considered in this work. The transition densities refer to low temperatures, $T\sim 1\,\mathrm{MeV}$.

\begin{table}[h]
    \centering
    \begin{tabular}{c|cccc}
    \hline
    \hline
    $L_{\rm sym}\,\rm [MeV]$ & $\eta_D$ & $\eta_V$ &
    $B^*\,\mathrm{[MeV/fm^3]}$ & $\varrho_{\rm tr}\,\mathrm{[fm^{-3}]}$ \\
    \hline
    \multicolumn{5}{c}{} \\[-0.8em]
    \multicolumn{5}{c}{\textbf{NJL(cutoff)}} \\
    \multicolumn{5}{c}{} \\[-0.5em]
    \hline
    50 & 1.0  & 0.8 & -15.0 & 0.55 \\
    50 & 1.0  & 1.0 & -20.0 & 0.57 \\
    50 & 1.0  & 1.2 & -23.0 & 0.61 \\
    50 & 1.25 & 0.8 & -5.0  & 0.47 \\
    50 & 1.25 & 1.0 & -7.0  & 0.50 \\
    50 & 1.25 & 1.2 & -15.0 & 0.52 \\
    \hline
    30 & 1.0  & 1.0 & -20.0 & 0.57 \\
    30 & 1.25 & 1.0 & -7.0  & 0.51 \\
    \hline
    70 & 1.0  & 1.0 & -20.0 & 0.57 \\
    70 & 1.25 & 1.0 & -7.0  & 0.49 \\
    \hline
    \multicolumn{5}{c}{} \\[-0.8em]
    \multicolumn{5}{c}{\textbf{NJL(RG)}} \\
    \multicolumn{5}{c}{} \\[-0.5em]
    \hline
    50 & 1.0  & 1.0 & -20.0 & 0.54 \\
    50 & 1.25 & 1.0 & -7.0  & 0.47 \\
    \hline
    \hline
    \end{tabular}
    \caption{Values of the bag constant $B^*$ and transition densities
    $\varrho_{\rm tr}$ for the different models considered in the work.}
    \label{tab:1}
\end{table}

\section{Numerical phase boundaries and maximum-mass configurations}
\label{chapter:addition}
For completeness, Table~\ref{tab:big_table} reports the onset and
completion baryon densities of the mixed phase,
$\varrho_{\rm I}$ and $\varrho_{\rm II}$, along with the maximum
mass and its radius for each indicated isentropic configuration.
These data supplement the parameter-dependent mass shifts presented
in Table~\ref{tab:masses} of the main text.

\begin{table*}
    \centering
    \begin{tabular}{lccccccc}
        \hline
        \hline
       Model  & $\eta_D$ &$\eta_V$ &$S$  &$\varrho_{\rm I}\,[  \rm fm^{-3}]$  &$\varrho_{\rm II}\,[  \rm fm^{-3}]$  & $M_{\rm max}[M_{\odot}]$  & $R_{\rm max}\,[\rm km]$ \\
 \hline
 \multicolumn{8}{c}{} \\[-0.8em]
    \multicolumn{8}{c}{\textbf{NJL(cutoff)}} \\
    \multicolumn{8}{c}{} \\[-0.5em]
    \hline
        LS30+NJL & 1.0 & 1.0 & 1 & 0.62 & 0.90 & 2.37 & 12.8 \\
        LS30+NJL & 1.0 & 1.0 & 2 & 0.56 & 0.85 & 2.37 & 13.6 \\
        LS30+NJL & 1.25 & 1.0 & 1 & 0.53 & 0.65 & 2.27 & 13.0 \\
        LS30+NJL & 1.25 & 1.0 & 2 & 0.49 & 0.63 & 2.27 & 14.0 \\
       \hline
        LS50+NJL & 1.0 & 0.8 & 1 & 0.59 & 0.83  & 2.33 & 12.9 \\
        LS50+NJL & 1.0 & 0.8 & 2 & 0.53 & 0.79  & 2.34 & 13.8 \\
        LS50+NJL & 1.0 & 1.0 & 1 & 0.62 & 0.90 & 2.36 & 12.8 \\
        LS50+NJL & 1.0 & 1.0 & 2 & 0.56 & 0.85 & 2.37 & 13.6 \\
        LS50+NJL & 1.0 & 1.2 & 1 & 0.66 & 0.98 & 2.39 & 12.6 \\
        LS50+NJL & 1.0 & 1.2 & 2 & 0.60 & 0.93 & 2.40 & 13.4 \\
        LS50+NJL & 1.25 & 0.8 & 1 & 0.49 & 0.59  & 2.16  & 12.9  \\
        LS50+NJL & 1.25 & 0.8 & 2 & 0.46 & 0.58  & 2.17 & 14.0 \\
        LS50+NJL & 1.25 & 1.0 & 1 & 0.53 & 0.65 & 2.27 & 13.0 \\
        LS50+NJL & 1.25 & 1.0 & 2 & 0.49 & 0.62 & 2.27 & 14.0 \\
        LS50+NJL & 1.25 & 1.2 & 1 & 0.56 & 0.68 & 2.31 & 12.9 \\
        LS50+NJL & 1.25 & 1.2 & 2 & 0.50 & 0.64 & 2.30 & 13.9 \\
        \hline
        LS70+NJL & 1.0 & 1.0 & 1 & 0.62 & 0.90 & 2.36 & 12.8 \\
        LS70+NJL & 1.0 & 1.0 & 2 & 0.58 & 0.85 & 2.37 & 13.6 \\
        LS70+NJL & 1.25 & 1.0 & 1 & 0.53 & 0.64 & 2.26 & 13.0 \\
        LS70+NJL & 1.25 & 1.0 & 2 & 0.48 & 0.62 & 2.27 & 14.0 \\
        
  \hline
   \multicolumn{8}{c}{} \\[-0.8em]
    \multicolumn{8}{c}{\textbf{NJL(RG)}} \\
    \multicolumn{8}{c}{} \\[-0.5em]
    \hline
    LS50+NJL(RG) & 1.0 & 1.0 & 1 & 0.58 & 0.80 & 2.32 & 13.0 \\
    LS50+NJL(RG) & 1.0 & 1.0 & 2 & 0.52 & 0.76 & 2.33 & 13.9 \\
    LS50+NJL(RG) & 1.25 & 1.0 & 1 & 0.48 & 0.57  & 2.16 & 13.0 \\
    LS50+NJL(RG) & 1.25 & 1.0 & 2 & 0.45 & 0.57 & 2.17 & 14.2 \\
  
  \hline
    \end{tabular}
    \caption{Mixed-phase onset and completion baryon densities
    $\varrho_{\rm I}$ and $\varrho_{\rm II}$, maximum gravitational
    masses, and corresponding radii of the isentropic configurations
    for the DDLS hadronic models (denoted LS30, LS50, LS70) and the
    NJL models with sharp-cutoff and RG-consistent regularization.
    All results refer to trapped matter at $Y_{Le}=0.4$, using the
    global-lepton-number (Gibbs) construction.}
    \label{tab:big_table}
\end{table*}

\clearpage
\bibliographystyle{apsrev4-2}
\bibliography{Ref,2sc_eos,refs_springer} 
\end{document}